%% file: paper.tex
\newcommand{\PaperTitle}{Every Keystroke You Make: A Tech-Law Measurement and Analysis of Event Listeners for Wiretapping}

\documentclass[letterpaper,twocolumn,10pt]{article}
\usepackage[table]{xcolor} 
\usepackage{usenix}

\usepackage{graphicx}
\usepackage{array}
\usepackage{xspace}
\usepackage{subcaption}
\usepackage{makecell}
\usepackage{placeins}
\usepackage{tikz}
\usetikzlibrary{arrows.meta,positioning,patterns,patterns.meta}
\usepackage{booktabs} 
\usepackage{pifont}
\usepackage[T1]{fontenc}
\usepackage{hyphenat}
\usepackage{enumitem}
\usepackage{mdframed}
\usepackage{fontawesome5}

\definecolor{lightgray}{gray}{0.85}
\global\mdfdefinestyle{insightstyle}{%
	outerlinewidth=0.5pt,
	backgroundcolor=gray!10,
	linecolor=gray,
	innertopmargin=6pt,
	innerbottommargin=6pt,
	innerrightmargin=7pt,
	innerleftmargin=6pt
}

\newenvironment{lesson}[1][Takeaways.]
{%
\smallskip
\begin{mdframed}[style=insightstyle]
\textbf{#1}
}
{%
\end{mdframed}
}

\newenvironment{infobox}[1][]
{%
\smallskip
\begin{mdframed}[style=insightstyle]
\textbf{#1}
}
{%
\end{mdframed}
}

\input{MacrosMarkup}

\input{macros}

\begin{document}

\title{\PaperTitle}

\author{
{\rm Shaoor Munir}\\
UC Davis \\
smunir@ucdavis.edu
\and
{\rm Nurullah Demir}\\
Institute for Internet-Security\\
demir@internet-sicherheit.de
\and
{\rm Qian Li}\\
Maastricht University\\
q.li@maastrichtuniversity.nl
\and
{\rm Konrad Kollnig}\\
Maastricht University\\
konrad.kollnig@maastrichtuniversity.nl
\and
{\rm Zubair Shafiq}\\
UC Davis\\
zubair@ucdavis.edu
} 

\maketitle

\label{sec:abstract}
\input{parts/abstract}

\section{Introduction}
\label{sec:introduction}
\input{parts/introduction}

\section{Background \& Related Work}
\label{sec:background}
\input{parts/background}

\section{Data Collection}
\label{sec:measurements}
\input{parts/measurements}


\section{Measurement \& Analysis}
\label{sec:wiretapping}
\input{parts/wiretappers}

\section{Legal Analysis and Scope}
\label{sec:legal_analysis}
\input{parts/legal_analysis}

\section{Conclusion}
\label{sec:discussion conclusions}
\input{parts/discussion}


\bibliographystyle{plain}
\bibliography{references}

\appendix
\input{parts/appendix}

\end{document}

%% file: MacrosMarkup.tex
\providecommand{\UsePackageFor}[2]{ \ifx#2\undefined\usepackage{#1}\fi }

\UsePackageFor{amssymb}{\boxtimes}
\usepackage[normalem]{ulem}		
\usepackage{bbding}				
\usepackage{endnotes}			
\usepackage{hyperref}			
\usepackage{hyperendnotes}	
\usepackage{pdfcolfoot}			

\ifdefined\OrigFootnote\relax\else
	\newenvironment{FootnoteContent}{}{}
	\let\OrigFootnote\footnote
	\let\OrigFootnoteText\footnotetext
	\renewcommand{\footnotetext}[1]{\OrigFootnoteText{\begin{FootnoteContent}#1\end{FootnoteContent}}}
	\renewcommand{\footnote    }[1]{\OrigFootnote    {\begin{FootnoteContent}#1\end{FootnoteContent}}}
\fi

\definecolor{PurplePlum}{rgb}{0.1,0,0.55} 
\definecolor{Brown}{rgb}{0.5,.25,0}
\definecolor{Orange}{rgb}{1,.3,0}
\definecolor{Gray}{rgb}{.7,.7,.7}
\definecolor{DarkGreen}{rgb}{.1,.41,.1}

\newif\ifBleck

\newcommand\Colour[1] {\color{#1}}

\newcommand\PrintToCLinks{	
  {\Colour{blue}\mbox{
    \hyperlink{w1619}{\sf$\rightarrow$~top}\quad
    \hyperlink{w1031}{\sf$\rightarrow$~toc}\quad
    \hyperlink{w1148}{\sf$\rightarrow$~lof}\quad
    \hyperlink{GreenRoom}{\sf$\rightarrow$~gr}\quad
    \hyperlink{EndNotes}{\sf$\rightarrow$~en}\quad
    \hyperlink{Sargasso}{\sf$\rightarrow$~sg}\quad
    \hyperlink{Index}{\sf$\rightarrow$~idx}
  }}
}
\makeatletter 
\newcommand\ToCLinks{
  \ifx\@onlypreamble\@notprerr		
    \hypertarget{w1619}{}			
  \else
    \AtBeginDocument{\hypertarget{w1619}{}}	
  \fi

  \ifBleck\else	
    \ifdefined\cofoot
      \cofoot{\PrintToCLinks}
      \cefoot{\PrintToCLinks}
    \else
      \def\@oddfoot{\PrintToCLinks}
      \def\@evenfoot{\PrintToCLinks}
    \fi
 \fi
}
\makeatother

\newif\ifEndNotes 

\newcommand\FnSym{{\scriptsize\PencilLeftDown\kern.1em}}		
\newcommand\EnSym {{$\bigtriangledown$}}

\def\MarkupsHowto{} 
\newcommand{\MarkupsHowtoAdd}[1]{\expandafter\def\expandafter\MarkupsHowto\expandafter{\MarkupsHowto{}#1}} 

\newif\ifMarkupsHowtoPrinted 
\newif\ifSuppress 

\newcommand\MakeMarkups[3][.]{

     \Suppressfalse
     \ifBleck\Suppresstrue\fi
     \ifx0#1\Suppresstrue\fi
     \ifx1#1\Suppressfalse\fi
     
     \expandafter\providecommand\csname#2x\endcsname {} 
     \ifSuppress\expandafter\renewcommand\csname#2x\endcsname{\relax}\else
                       \expandafter\renewcommand\csname#2x\endcsname{#3}\fi
                       
     \expandafter\providecommand\csname#2\endcsname {} 
     \ifSuppress\expandafter\renewcommand\csname#2\endcsname[1]{##1}\else
                       \expandafter\renewcommand\csname#2\endcsname[1]{{\csname#2x\endcsname##1}}\fi

     \expandafter\providecommand\csname#2d\endcsname {} 
     \ifSuppress\expandafter\renewcommand\csname#2d\endcsname[1]{\relax}\else
                       \expandafter\renewcommand\csname#2d\endcsname[1]{{\csname#2x\endcsname\sout{##1}}}\fi
                       
     \expandafter\providecommand\csname#2r\endcsname {} 
     \ifSuppress\expandafter\renewcommand\csname#2r\endcsname[2]{{##2}}\else
                       \expandafter\renewcommand\csname#2r\endcsname[2]{\csname#2d\endcsname{##1} \csname#2\endcsname{##2}}\fi

     \expandafter\providecommand\csname#2i\endcsname {} 
     \ifSuppress\expandafter\renewcommand\csname#2i\endcsname[1]{\relax}\else
                       \expandafter\renewcommand\csname#2i\endcsname[1]{\csname#2\endcsname{##1}}\fi

     \expandafter\providecommand\csname#2t\endcsname {} 
     \ifSuppress\expandafter\renewcommand\csname#2t\endcsname[1]{\relax}\else
                       \expandafter\renewcommand\csname#2t\endcsname[1]{{\csname#2x\endcsname{\mbox{$\langle\!\langle$}##1{\csname#2x\endcsname\mbox{$\rangle\!\rangle$}}}}}\fi 

     \expandafter\providecommand\csname#2b\endcsname {} 
     \ifSuppress\expandafter\renewcommand\csname#2b\endcsname[1][empty]{\relax}\else 
                       \expandafter\renewcommand\csname#2b\endcsname[1][\empty]{\ifx\empty##1\empty
                       	\label{#2-bookmark} 
                              \marginpar [\raggedleft\csname#2\endcsname{{\footnotesize\fbox{#2 working here}}~$\Longrightarrow$}]
                                                {\csname#2\endcsname{$\Longleftarrow$~{\footnotesize\fbox{#2 working here}}}}
                       \else 
                       	\marginpar [\raggedleft\csname#2\endcsname{\ifx\empty##1\empty\else\fbox{\tiny\parbox{8em}{\raggedright##1}}~\fi$\Longrightarrow$}]
                                                {\csname#2\endcsname{$\Longleftarrow$\ifx\empty##1\empty\else~{\tiny\fbox{\parbox{8em}{\raggedright##1}}}\fi}}\fi}\fi

     \expandafter\providecommand\csname#2TD\endcsname {} 
     \ifSuppress\expandafter\renewcommand\csname#2TD\endcsname{\relax}\else
                       \expandafter\renewcommand\csname#2TD\endcsname{\csname#2\endcsname{\fbox{#2 to do}}}\fi

     \expandafter\providecommand\csname#2Bar\endcsname {} 
     \ifSuppress\expandafter\renewcommand\csname#2Bar\endcsname{\relax}\else
                       \expandafter\renewcommand\csname#2Bar\endcsname{\csname#2\endcsname{\scriptsize\XSolidBrush}}\fi

     \expandafter\providecommand\csname#2f\endcsname {} 
     \ifSuppress\expandafter\renewcommand\csname#2f\endcsname[2][]{\relax}\else
      \expandafter\renewcommand\csname#2f\endcsname[2][\empty]{ 
        {\mbox{\csname#2x\endcsname\tiny$\boxtimes$}\marginpar{\hsize1cm\csname#2x\endcsname\fbox{\FnSym\footnotemark}}\relax 
        \footnotetext{\csname#2x\endcsname##2}}}\fi

     \expandafter\providecommand\csname#2e\endcsname {}
     \ifSuppress\expandafter\renewcommand\csname#2e\endcsname[1]{\relax}\else%
      \expandafter\renewcommand\csname#2e\endcsname[1]{%
       \global\EndNotestrue
       \mbox{\scriptsize\csname#2x\endcsname$\boxtimes$}\relax%
       \marginpar{\hsize1cm\csname#2x\endcsname\fbox{\EnSym\endnotemark%
                          \hypertarget{ENmark\thepage-\theendnote}{}~\hyperlink{ENtext\thepage-\theendnote}{{\Colour{blue}$\downarrow$}}}%
       }%
       {
        \def\zz{\noexpand#3}%
        \edef\z{~{[Endnote \theendnote\ %
        on p.\noexpand\hypertarget{ENtext\thepage-\theendnote}{}\thepage%
                    ~\noexpand\hyperlink{ENmark\thepage-\theendnote}{{\noexpand\Colour{blue}$\uparrow$}}]}%
        }%
        \expandafter\endnotetext\expandafter{\z\vspace{2ex}\\ ##1\newpage}%
       }
      }\fi

     \expandafter\providecommand\csname#2n\endcsname {}
     \ifSuppress\expandafter\renewcommand\csname#2n\endcsname[1]{\relax}\else%
      \expandafter\renewcommand\csname#2n\endcsname[1]{%
       \global\EndNotestrue
    \marginpar{{\tiny\endnotemark}\hypertarget{ENmark\thepage-\theendnote}{}~\hyperlink{ENtext\thepage-\theendnote}{}}
       {
        \def\zz{\noexpand#3}%
        \edef\z{~{\zz[Endnote (deferred) 
        from p.\noexpand\hypertarget{ENtext\thepage-\theendnote}{}\thepage%
        ]}%
        }%
        \expandafter\endnotetext\expandafter{\z\vspace{2ex}\\ ##1\newpage}%
       }
      }\fi

     \expandafter\providecommand\csname#2fe\endcsname {} 
     \ifSuppress\expandafter\renewcommand\csname#2fe\endcsname[2][]{\relax}\else 
      \expandafter\renewcommand\csname#2fe\endcsname[2][]{ 
       \def\File{##1}\relax
       \ifx\File\empty\csname#2f\endcsname{##2}\else 
        \global\EndNotestrue 
        \mbox{\scriptsize\csname#2x\endcsname$\boxtimes$}
        \marginpar{\csname#2x\endcsname\fbox{\FnSym\footnotemark}}\relax
        \footnotetext{~\csname#2x\endcsname##2\
                             --- See [\EnSym\endnotemark\hypertarget{ENmark\thepage-\theendnote}{}
                             \kern-.2em\hyperlink{ENtext\thepage-\theendnote}{{\Colour{blue}$\downarrow$}}].}\relax
       { 
         \def\zz{\noexpand#3}
         \edef\z{~{\zz[Endnote~\thefootnote~on~p.\noexpand\hypertarget{ENtext\thepage-\theendnote}{}\thepage
                     ~\noexpand\hyperlink{ENmark\thepage-\theendnote}
                     {{\noexpand\Colour{blue}\kern-0.1em$\uparrow$}]}}
                     {\noexpand\footnotesize\noexpand\newline\noexpand\hspace*{2em} (~from file {\noexpand\tt\File.tex}~)}
         }    
         \expandafter\endnotetext\expandafter{\z~\par\input{##1}\newpage}
        } 
       \fi 
      } 
     \fi 

     \ifSuppress\relax\else\ifBleck\relax\else
      \MarkupsHowtoAdd{\par\csname#2t\endcsname{
       $\backslash$\texttt{#2}$\cdots$\ markups are in \textbf{this} colour\ifx#1..\else\ifx1#1.\else, e.g.\ for #1.\fi\fi
       \ifMarkupsHowtoPrinted\relax\else 
        \global\MarkupsHowtoPrintedtrue 
        \begin{quote}\begin{tabular}{l@{\hspace{2em}}p{.7\linewidth}}
         \multicolumn{2}{l}{\texttt{$\backslash$MakeMarkups\ifx#1.\relax\else[#1]\fi\{#2\}\{{\it$\langle$colour command\/$\rangle$}\}}
         				 --- Defines the macros below:}\\
             & see comments at \texttt{$\backslash$MakeMarkups} definition. \\[1ex]
         \texttt{$\backslash$#2\{$\langle$text$\rangle$\}} & Sets \texttt{$\langle$text$\rangle$} in \texttt{#2}'s colour. \\
         \texttt{$\backslash$#2x} & Changes to \texttt{#2}'s colour (until end of context). \\
         \texttt{$\backslash$#2d\{$\langle$text$\rangle$\}} & Sets \texttt{$\langle$text$\rangle$} in \texttt{#2}'s colour with a strikethrough (i.e.\ delete). \\
         \texttt{$\backslash$#2r\{$\langle$this$\rangle$\}\{$\langle$that$\rangle$\}} &
          Strikes through \texttt{$\langle$this$\rangle$} and inserts \texttt{$\langle$that$\rangle$} (i.e.\ replace). \\
         \texttt{$\backslash$#2f\{$\langle$text$\rangle$\}} & Meta-comment: puts \texttt{$\langle$text$\rangle$} in a \texttt{#2}-footnote with a {\tiny$\boxtimes$} in the main text. \\
         \texttt{$\backslash$#2t\{$\langle$text$\rangle$\}} & Use for meta when  \texttt{$\backslash$#2f} isn't allowed (``Not in outer-par mode.'') \\
         \texttt{$\backslash$#2b[$\langle$optional$\rangle$]} & Marginal pointer, with label for hyper-linking directly there. \\
         \texttt{$\backslash$#2e\{$\langle$text$\rangle$\}} & Puts \texttt{$\langle$text$\rangle$} in a \texttt{#2}-endnote with a (big) $\boxtimes$ in the main text. \\[.5ex]
         \texttt{$\backslash$#2n\{$\langle$text$\rangle$\}} & Like \texttt{$\backslash$#2e}
         except there's no reference from the main text. Good for ``decluttering''
         when you still want to have the footnote- or endnote texts as reminders. \\[.5ex]
         \texttt{$\backslash$#2fe[$\langle$this$\rangle$]\{$\langle$that$\rangle$\}} & Makes a \texttt{$\backslash$#2f\{$\langle$that$\rangle$\}} that refers to a \\
           & \texttt{$\backslash$#2e\{$\langle$contents of file this.tex$\rangle$\}}. \\ 
           & Without the optional argument, acts as \texttt{$\backslash$#2f\{$\langle$that$\rangle$\}}. \\[.5ex]
         \texttt{$\backslash$#2Bar} & Inserts ``burn after reading'' symbol \csname#2Bar\endcsname, meaning
          \begin{quote}\begin{itemize}\setlength\itemsep{0pt}
           \item If yours is the only \csname#2Bar\endcsname\ in this (presumably someone else's) footnote, and you are happy that the footnote has been addressed,
           go ahead and comment-out the whole footnote. (The \csname#2Bar\endcsname\ is their request for you to ``approve and remove''.)
           \item If you are not happy, delete only your \csname#2Bar\endcsname\ and follow-on in the footnote
            (in your colour, i.e.\ with \texttt{$\backslash$#2x}) saying why you are not happy.
           \item If you are happy, but there are others' burn-after-reading symbols as well as yours, just delete yours; the other people have not yet responded.
          \end{itemize}
          \end{quote}
          The idea is that when everyone's happy, the last person will comment-out the meta-text. \\[0.5ex]
         \texttt{$\backslash$#2TD} & Inserts {\csname#2TD\endcsname}\ . \\
        \end{tabular}\end{quote}
       \fi
      }}
     \fi\fi
}

\newif\ifNoGreenRoom
\newcommand\MakeGreenRoom {\ifBleck\relax\else\ifNoGreenRoom\relax\else
\newcommand\NewGRLabel[1] {\OldGRLabel{GreenRoom-##1}} 
 \newcommand\NewGRRef[1] 
 {\expandafter\ifx\csname r@GreenRoom-##1\endcsname\relax\OldGRRef{##1}\else\OldGRRef{GreenRoom-##1}\fi}
 \let\OldGRLabel\label \let\label\NewGRLabel
 \let\OldGRRef\ref \let\ref\NewGRRef
 \hrule
 ~\\\begin{center}\Huge \hypertarget{GreenRoom}{Green Room}
 \end{center}~\\
 \hrule
\fi\fi}

\newcommand\EndGreenRoom  {\ifBleck\relax\else\ifNoGreenRoom\relax\else
\let\label\OldGRLabel
\let\ref\OldGRRef
\fi\fi}

\newif\ifNoEndNotes

\newif\ifNoSargasso
\newcommand\MakeSargasso {
 \hypertarget{Sargasso}{}
 \newcommand\NewLabel[1] {\OldLabel{Sargasso-##1}} 
 \newcommand\NewRef[1] 
 {\expandafter\ifx\csname r@Sargasso-##1\endcsname\relax\OldRef{##1}\else\OldRef{Sargasso-##1}\fi}
 \let\OldLabel\label \let\label\NewLabel
 \let\OldRef\ref \let\ref\NewRef
\ifBleck\end{document}\else\ifNoSargasso
\relax
\else
  \hrule
  ~\\\begin{center}\Huge Sargasso
  \end{center}~\\
  \hrule
 \fi\fi
}

\newcommand\EndSargasso  {\ifBleck\relax\else\ifNoSargasso\relax\else
\let\label\OldLabel
\let\ref\OldRef
\fi\fi}

\newcommand\EndDocument {\ifBleck\end{document}\fi} 

\newcommand\Cite[2][\empty] {{\Colour{red}\ifx#1\empty[#2]\else[#2,~#1]\fi}}



%% file: macros.tex
\MakeMarkups[Shaoor]{S}{\Colour{blue}}
\MakeMarkups[Zubair]{Z}{\Colour{orange}}
\MakeMarkups[Konrad]{K}{\Colour{teal}}
\MakeMarkups[Qian Li]{Q}{\Colour{brown}}
\MakeMarkups[Nurullah]{N}{\Colour{red}}

\newcommand{\para}[1]{\addvspace{0.5\baselineskip}\noindent\textbf{#1}}

\newcommand{\nsites}{15k\xspace}

\newcommand{\navgeventlisteners}{48.8\xspace}
\newcommand{\npcteventlisteners}{91.48\%\xspace}

\newcommand{\nwtapping}{3.18\%\xspace}
\newcommand{\nweventlisteners}{38.52\%\xspace}

\newcommand{\listenersthirdparty}{81.3\%\xspace}

\newcommand{\wiretapping}{wiretapping\xspace}
\newcommand{\Wiretapping}{Wiretapping\xspace}
\newcommand{\wiretap}{wiretap\xspace}

\newcommand{\wiretapped}{wiretapped\xspace}
\newcommand{\Wiretapped}{Wiretapped\xspace}
\newcommand{\wiretapper}{wiretapper\xspace}

\newcommand{\wiretappers}{wiretappers\xspace}

\newcommand{\eg}{e.g.,\xspace}

\newcommand{\tikzxmark}{%
\tikz[scale=0.23] {
    \draw[line width=0.7,line cap=round, color=red!100!black] (0,0) to [bend left=6] (1,1);
    \draw[line width=0.7,line cap=round, color=red!90!black] (0.2,0.95) to [bend right=3] (0.8,0.05);
}}
\newcommand{\tikzcmark}{%
\tikz[scale=0.23] {
    \draw[line width=0.7,line cap=round, color=green!80!black] (0.25,0) to [bend left=10] (1,1);
    \draw[line width=0.8,line cap=round, color=green!80!black] (0,0.35) to [bend right=1] (0.23,0);
}}


%% file: parts/abstract.tex
\begin{abstract}
The privacy community has a long track record of investigating emerging types of web tracking techniques. 
Recent work has focused on compliance of web trackers with new privacy laws such as Europe's GDPR and California's CCPA. 
Despite the growing body of research documenting widespread lack of compliance with new privacy laws, there is a lack of robust enforcement. 
Different from prior work, we conduct a tech-law analysis to map decades-old U.S. laws about interception of electronic communications---so-called \textit{wiretapping}---to web tracking. 
Bridging the tech-law gap for older wiretapping laws is important and timely because, in cases where legal harm to privacy is proven, they can provide statutory private right of action, are at the forefront of recent privacy enforcement, and could ultimately lead to a meaningful change in the web tracking landscape.

In this paper, we focus on a particularly invasive tracking technique: the use of JavaScript event listeners by third-party trackers for real-time keystroke interception on websites.
%
%
We use an instrumented web browser to crawl a sample of the top-million websites to investigate the use of event listeners that aligns with the criteria for wiretapping, according to U.S. wiretapping law at the federal level and in California. 
We find evidence that \nweventlisteners websites installed third-party event listeners to intercept keystrokes, and that \textit{at least} \nwtapping websites transmitted intercepted information to a third-party server, which aligns with the criteria for wiretapping. 
We further find evidence that the intercepted information such as email addresses typed into form fields (even without form submission) are used for unsolicited email marketing.
Beyond our work that maps the intersection between technical measurement and U.S. wiretapping law, additional future \textit{legal} research is required to determine when the wiretapping observed in our paper passes the threshold for illegality.
\end{abstract}

%% file: parts/introduction.tex
The modern web has come to rely on a complex and ubiquitous tracking ecosystem.
Websites deploy third-party tracking scripts to monitor user interactions—such as typing, clicking, and scrolling—often in real time and without the user's awareness, using the captured data for analytics and targeted advertising.
Determining when such web tracking crosses the line into unauthorized interception is important for both technical researchers and legal scholars, as it informs the applicability of relevant laws and regulations.

Legal protections against interception long pre-date the Internet. 
Federal and state ``wiretapping'' laws in the U.S.---most notably the federal Electronic Communications Privacy Act (ECPA, 1986) \cite{ecpa1986} and the California Invasion of Privacy Act (CIPA, 1967) \cite{california_penal_code_631}---were enacted to protect the contents of private communications from unauthorized interception.
Although initially focused on telephone lines, courts have since extended these statutes to new technologies, including e-mail \cite{USvCouncilman}, third-party cookies and pixels \cite{facebookinternettrackinglitigation,in_re_facebook_2017}, and most recently session replay scripts \cite{Javier_v_ASSURANCE_IQ_LLC,javier_assurance_2020}.
Each extension requires a careful mapping between the legal text and the technical specifics of the interception technique.
This interpretive exercise becomes increasingly challenging as web tracking methods grow more complex.

The research community has built a rich body of empirical work on measuring web tracking (e.g., \cite{mayer2012third,acar2014web,nikiforakis2013cookieless,acar2013fpdetective,bashir2016tracing,Laperdrix2016BeautyBeast,Englehardt2016OnlineTracking,acar2020no,dao2021cname,Siby2022WebGraph,Munir2023CookieGraph,lin2023fashion,Munir2024PURL,fouad2024devil}).
Recent studies have tried to bridge the tech-law gap emerging from novel privacy laws such as the European Union GDPR (2016) and the Californian CCPA/CPRA (2018/2020), focusing primarily on the issue of \textit{consent}: whether cookie banners comply with legal requirements and whether users meaningfully engage with them \cite{Degeling.2018,Utz.2019,vanhong.2024.compliance,bielova2024effect,matte2020cookie,kollnig_consent,nouwens_consent}.
Other work probes sector-specific contexts---health \cite{Wesselkamp2021ERNIE,Huo2022AllEyesOnMe,Yu2022GotSickTracked} and child-directed content \cite{Moti2024TargetedTroublesome,Reyes2018COPPACompliance,Vlajic2018OnlineTracking}---where U.S. laws like HIPAA and COPPA apply.

While there is prior research on the prevalence of web tracking and different types of tracking techniques (e.g., cookies, fingerprinting), there is a lack of prior research on whether web tracking, as widely studied by the research community, fits the definition of wiretapping.
There are several reasons why this line of work is important and timely.
First, unlike sector-specific privacy laws in the U.S. \cite{epic_us_privacy_laws}, wiretapping laws are not limited to communication on health-related (as is HIPAA) or child-directed (as is COPPA) websites but are  applicable to any type of communication.
Hence, these wiretapping laws have the potential to offer broad privacy protections.
Second, wiretapping laws in the U.S. provide statutory private right of action allowing consumers to obtain injunctive and monetary relief.
Hence, unlike other privacy laws in the U.S. such as CCPA/CPRA, HIPAA, and COPPA, which have no or limited private right of action and thus have not been rigorously enforced, wiretapping laws allow consumers to bring collective civil enforcement actions \cite{yang2024state}.
Third, there have been struggles with the enforcement of data protection laws in Europe, particularly the GDPR~\cite{lynskey_grappling_2019,masse_two_2020,iccl_ads}.
While wiretapping laws exist in individual European countries, they do not provide private right of action and no wiretapping-specific law exists at a European level~\cite{europaFrequentlyAsked}.
This also makes the focus on the wiretapping laws in the U.S. in this paper important and timely.

In order to apply CIPA’s (California's wiretapping statute) definition of “wiretapping” to today’s web ecosystem, we focus on a specific mechanism that third-party trackers use to intercept user inputs: JavaScript \emph{event listeners}.
Different from prior work on web tracking, this requires an understanding of \textbf{whether} and \textbf{how} contents of communication between Alice (consumer) and Bob (website) are intercepted by Eve (third-party tracker). 
We first analyze the statutory elements of wiretapping under CIPA and translate them into concrete, testable technical criteria.
Using those criteria, we perform the first large-scale measurement of how event listeners are installed, invoked, and subsequently used to transmit intercepted keystrokes or form inputs to remote servers.

\textbf{Note that our study aims to measure the extent of potential wiretapping on the web, focusing solely on detecting instances where keystrokes are intercepted in real-time and transmitted to remote servers.
Determinations regarding the legality of specific practices---including consent and other factual nuances---are beyond the scope of this work and require further technical and legal analysis. We provide a detailed discussion in Appendix} \ref{app:legal_analysis}.

\para{This paper makes the following contributions:}

\para{As a technical contribution,} we study the use of event listeners for web tracking, particularly the secret interception and collection of users' keystrokes and form inputs.
To this end, we develop an instrumented web browser to measure the use of event listeners on websites.
Computationally, the measurement of event listeners is not trivial.
This is because a single webpage may have hundreds of event listeners installed by different third-party scripts and attached to various DOM elements, all of which would need to be monitored for both installation and invocation in the analysis of a single website.
We use our instrumented web browser to study the use of event listeners on a \nsites sample of the top-million websites.

\para{As a legal contribution,} with the help of legal experts, we investigate when the use of event listeners by trackers might align with the criteria for wiretapping under California's CIPA.
Originally, wiretapping laws were enacted to protect communications on telephone networks~\cite{wiretap1968,california_penal_code_631} but have since been interpreted to extend to internet communications. 
Recent civil enforcement actions have raised allegations that certain web trackers, particularly session replay scripts, may violate these statutes~\cite{Saleh_v_Nike, Javier_v_ASSURANCE_IQ_LLC, Popa_v_Harriet_Carter_Gifts}.
Rather than making legal determinations, our work aims to provide empirical insights to inform ongoing discussions about the potential intersection of web tracking and wiretapping.

\para{Key Findings.}
Our measurement and analysis reveal widespread use of JavaScript event listeners on \npcteventlisteners of the tested websites, with \navgeventlisteners event listeners installed on average per website.
We find that most event listeners are set by third-party scripts, as much as \listenersthirdparty. 
Among these, we identify \texttt{keydown}, \texttt{keyup}, and \texttt{keypress} as capable of directly intercepting user keystrokes, capturing the content of user communications.
By analyzing onward sharing of the intercepted user keystrokes with remote servers, we provide evidence of wiretapping happening on \textit{at least} \nwtapping of the tested websites.
We further find evidence that intercepted user keystrokes such as email addresses typed into form fields (even without form submission) are used for email marketing.


\para{Scope of our tech-law analysis:} Our work aims to provide a lower bound of what kind of web tracking may fit the definition of wiretapping. 
Importantly, beyond keystroke interception with event listeners as studied in this paper, there likely exist other types of web tracking that may also fit the definition of wiretapping.
At the same time, it is important to note that not all wiretapping is illegal. Such an assessment depends on additional factual and legal discussion, potentially in courts, that are beyond the scope of this paper.

The rest of the paper proceeds as follows. In Section \ref{sec:background}, we provide a technical background of event listeners and web tracking as well as describe historical evolution of wiretapping laws in the U.S.
We also present our tech-law threat model of wiretapping.
Section \ref{sec:measurements} describes our browser instrumentation and data collection.
In Section \ref{sec:wiretapping}, we present our measurements and analysis of event listeners in the wild. 
We report on the nature and prevalence of tracking scripts that fit our derived technical criteria.
We discuss our work's scope and limitations in Section \ref{sec:legal_analysis}.
We conclude with a discussion of potential implications of our work in Section \ref{sec:discussion conclusions}.

%% file: parts/background.tex
In this section, we provide technical background about web tracking as well as the use of event listeners for web tracking.
We then provide a legal analysis of U.S. wiretapping laws and the current interpretation of those laws by U.S. courts. 

\subsection{Technical Literature Review of Web Tracking and Event Listeners}
\label{subsec:event-listeners}
\para{Overview of Web Tracking and Event Listeners.} The research community has a long track record of investigating emerging web tracking techniques and developing technical countermeasures (e.g., \cite{mayer2012third,acar2014web,nikiforakis2013cookieless,acar2013fpdetective,bashir2016tracing,Laperdrix2016BeautyBeast, Englehardt2016OnlineTracking, acar2020no, dao2021cname, Siby2022WebGraph, Munir2023CookieGraph, lin2023fashion, Munir2024PURL,fouad2024devil}). 
At a high level, web tracking techniques can be divided into two categories: (1) techniques to identify users across websites and (2) techniques to collect the contents of communication between the user and the website.
Much of the prior research is focused on the former---tracking techniques such as cookies and fingerprinting that aim to identify users on and across websites. 
There is some research focused on tracking techniques such as session replay that aim to collect information about how a user interacts on a webpage (e.g., load a page, enter an email in a form field, or click on a button) \cite{Starov2016AreYouSure,acar2020no,Senol2022LeakyformsUSENIX}.
Our focus in this work is on the latter.

At the most basic level, a tracker that aims to collect information about user communication on a webpage can collect the domain or the URL of the webpage visited by a user. 
This can be accomplished in several ways. 
For example, a tracker, even if it is implemented as an image pixel, can rely on the HTTP Referer header \cite{MDNRefererHeader} in the request to its server. 
A tracking script (tracker that is implemented via JavaScript) can further make use of various powerful Web APIs to this end \cite{MDNWebAPIs}. 
For example, a tracker can use \texttt{window.location} \cite{MDNWindowLocation} or \texttt{document.URL} \cite{MDNDocumentURL}. 
Beyond the domain or the URL of a webpage, the Web APIs provide additional capabilities that can be used by a tracking script to collect additional information about a user's activity on the webpage. 
For example, a tracking script can access the HTML DOM---which is a representation of the structure and content of the HTML document---through a number of Web APIs \cite{MDNDOM}.
Through the \textit{Document} interface \cite{MDNDocumentInterface}, a tracking script can access contents of specific elements using methods such as \texttt{document.getElementById} and \texttt{document.getElementsByName}.

The Web APIs further allow scripts to collect fine-grained and real-time information about specific events happening during the loading of a webpage. 
For example, a script can use the \texttt{event} interface \cite{MDNEventInterface} to capture information about a wide range of page load events, such as the \texttt{load} event that indicates when the webpage has finished loading \cite{MDNWindowLoadEvent} as well as user actions events such as key presses via the \texttt{KeyboardEvent} interface \cite{MDNKeyboardEvent} and mouse movements via the \texttt{MouseEvent} interface \cite{MDNMouseEvent}. 

To use events for finer‑grained, real‑time tracking, a tracking script first installs an event listener that is configured to listen to a specific event such as \texttt{keydown} \cite{MDNElementKeydownEvent}.
Because the listener is registered directly on the element (or, in the case of delegated listeners, on an ancestor), it sits \emph{in‑path}: the browser dispatches the event to the listener before it bubbles up to parent elements, and the listener can even call \texttt{stopPropagation()} to halt further propagation.
This in‑path placement contrasts with classical \textit{on‑path} network traffic monitoring; here, the interception happens inside the execution context of the page before the page itself handles the event.

Real‑time interception is also \emph{possible} without event listeners---for instance, by polling DOM attributes or using Mutation Observers---but trackers rely on event listeners because they offer precise, low‑latency avenue for interception.
It is important to note that not every use of event listeners constitutes wiretapping.
For example, analytics frameworks often attach benign \texttt{click} listeners to measure aggregate engagement metrics, whereas the threat model we analyze targets listeners that capture user‑generated content (e.g., keystrokes in password or search fields) and transmit it to a remote endpoint before the page or user can meaningfully react.

When the monitored event occurs, JavaScript execution is directed to the event‑handler function of the registered listener \cite{MDNEventHandlers}.%
The handler can then immediately share the event information (e.g., which key was pressed) to a remote server via an XHR or \texttt{fetch} request \cite{MDNXMLHttpRequest}.
%


\para{Related Work on Event Listeners and Session Replay. }
Prior work has shown that event listeners are extensively used to track user activity on a webpage.
\cite{Atterer2006KnowingWWW} described how standard web technologies such as event listeners can enable collection of detailed user interactions on a webpage such as mouse movements, scroll, mouse click and over, and key press information including the key that was pressed.
The authors describe how this data can be used to infer high-level behaviors such as user hesitation in filling a form or usability testing.
Using this approach, prior work has built tools and systems for user session replay that can be used for web application testing \cite{Elbaum2003ImprovingWeb,Sprenkle2005AutomatedReplay}.

Jang et. al~\cite{Jang2010PrivacyViolatingIFCCS} conducted an empirical study of privacy-violating information flows in JavaScript, including the use of event listeners to track user behavior such as mouse and keyboard activity.
The authors found that event listeners were used to send keyboard input and mouse clicks of users on a quarter of the top-1,300 tested sites.
More importantly, the authors found that event listeners were used to perform covert tracking of keyboard and mouse activity, i.e. the user would be unaware that their information is being tracked by any other party on the webpage, via tracking libraries.

Researchers have also investigated the use of session replay scripts by third-party tracking scripts.
\cite{Englehardt2017NoBoundaries} found that third-party session replay scripts record and share sensitive medical and financial information as well as other personally identifiable information with  third-parties.
The authors found that while some session replay script providers try to redact sensitive data such as passwords, imperfect implementation of this on both the session replay provider's and website developers' end resulted in sensitive information leaks.
\cite{Senol2022LeakyformsUSENIX} also detected the collection and exfiltration of user-entered passwords through session replay services on more than 50 websites. 
\cite{Yu2022GotSickTracked} also detected the use of session replay services on sensitive websites that contain medical information.
Their analysis shows session replay scripts by Hotjar, Yandex, and FullStory were collecting personal information such as email, phone numbers, and user chats on hospital websites.

Grodzinsky et. al~\cite{Grodzinsky2022SessionreplayscriptTIS} further explored the privacy implications of using session replay scripts record user interactions on web applications.
They concluded that the data collected through session replay scripts often exceeds the intended usability improvements.
They also found that this data collection occurs without user consent and awareness.

While prior work has reported use of event listeners for accessing information entered by a user on a webpage, it lacks a systematic measurement of how event listeners are installed, invoked, and then subsequently the data collected through event listeners is transmitted to remote servers.
As we discuss later in this section, this particular use of event listeners can be likened to \wiretapping under certain state statutes in the United States.

\input{parts/law}

%% file: parts/law.tex
\subsection{Legal Literature  Review of Wiretapping and Application to Web Tracking}
\label{subsec:wiretapping}
\begin{figure*}[!htpb]
    \centering
    \begin{minipage}[t]{0.49\textwidth}
        \centering
        \includegraphics[width=\textwidth]{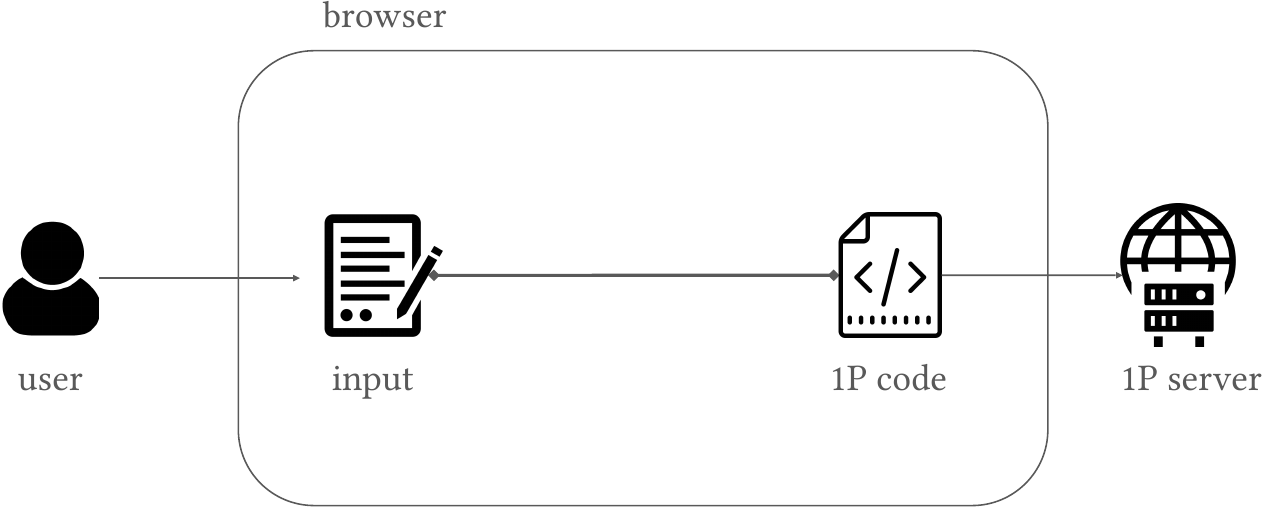}
        \subcaption{Normal input processing on a website.}
        \label{fig:input-event-listener-before}
    \end{minipage}
    \hfill
    \begin{minipage}[t]{0.49\textwidth}
        \centering
        \includegraphics[width=\textwidth]{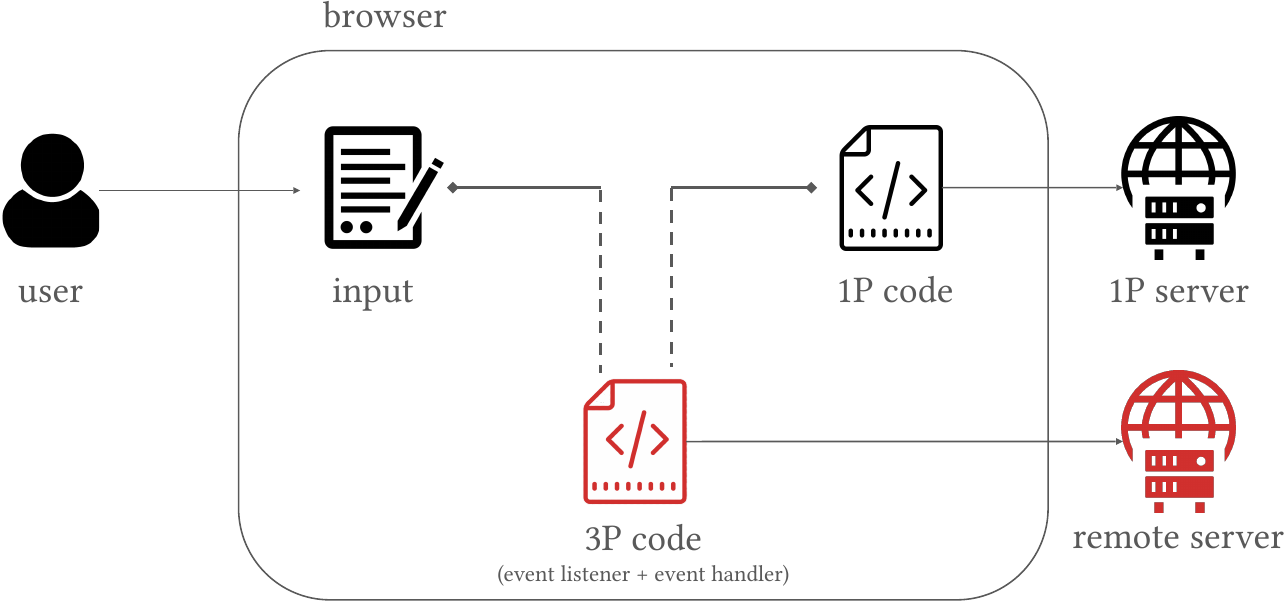}
        \subcaption{Input processing with a third-party event listener.}
        \label{fig:input-event-listener-after}
    \end{minipage}
    \caption{Comparison of input processing on a webpage with and without third-party event listeners.}
    \label{fig:input-event-listener}
\end{figure*}

\para{Legal History of U.S. Wiretapping Laws.} Wired communication became widespread following the invention of the telegraph and telephone in the 19th century.
This increasingly made wiretapping a powerful surveillance tool for authorities conducting criminal investigations.
Yet, it remained largely unregulated in the early 20th century, as demonstrated in the landmark 1928 decision \textit{Olmstead v.\ United States}~\cite{olmstead_v_us_1928}.
In this case, the U.S.\ Supreme Court ruled that wiretapping was permissible without a warrant because the government had not physically trespassed on Olmstead's property and did not need a search warrant under the Fourth Amendment~\cite{constitution_amendment_4}, except in exceptional circumstances~\cite{us_courts_educational_resources}.

To limit the reach of law enforcement following the \textit{Olmstead} decision, Congress enacted the first federal wiretapping law in 1934, the Communications Act~\cite{bja_privacy_civil_liberties}.
This law prohibited unauthorized wiretapping and made any evidence generated by it inadmissible in court.
Challenges arose as an increasing portion of communication, which also included criminal activity, took place over the wire~\cite{Kaplan2012HistoryWiretapping}.
The U.S.\ Supreme Court later ruled in the 1967 decision \textit{Katz v.\ United States}~\cite{constitution_center_katz} that installing a listening device without a warrant violated the Fourth Amendment, recognizing that conversations are protected as long as there is a “reasonable expectation of privacy,” regardless of physical intrusion~\cite{Kaplan2012HistoryWiretapping}.
In response, Congress enacted the Omnibus Crime Control and Safe Streets Act in 1968~\cite{wiretap1968} and its Title III---known as the \textit{Wiretap Act}.
This act codified the \textit{Katz} ruling by requiring law enforcement to obtain a warrant based on probable cause before conducting wiretapping~\cite{TitleIIIWiretapping}, thus creating a differentiation between \textit{lawful} and \textit{unlawful} wiretapping.

\para{Wiretapping under the Electronic Communications Privacy Act (ECPA).} With the enactment of the Electronic Communications Privacy Act (ECPA) in 1986~\cite{epic_ecpa}, Congress amended the Wiretap Act to extend its protections to electronic communications (e.g. email and fax).
The Act explicitly imposes liability on anyone who “intentionally intercepts, attempts to intercept, or directs another person to intercept or attempt to intercept any wire, oral, or electronic communication”~\cite{usc_2511_2520} and similarly applies to those who intentionally “disclose” or “use” a communication that has been unlawfully intercepted~\cite{usc_2511_c_d}.
Notably, ECPA authorizes interception when one party to the communication consents, \emph{unless} the interception is for the purpose of committing a criminal or tortious act~\cite{usc_2511_2_d}.
Because ECPA was enacted before the advent of modern electronic communication technologies (web, smartphones, and social media), courts continue to grapple with its application to emerging digital forms of communication.

\para{Wiretapping under the California Invasion of Privacy Act (CIPA).}
California’s Invasion of Privacy Act (CIPA), enacted in 1967, imposes a stricter consent requirement than the federal ECPA.
Unlike ECPA, which permits a single party to consent, CIPA requires \emph{all-party} (often called “two-party”) consent:
every participant to the communication must agree before wiretapping.
Section 631 of CIPA functions as California's wiretapping statute, making it unlawful, “without the consent of all parties to the communication,” to \emph{tap} or make an unauthorized connection; to willfully \emph{read}, or \emph{attempt to read or learn} the contents of a communication in transit; to \emph{use} or \emph{attempt to use} information so obtained; or to \emph{aid} or \emph{procure} any such act.

In later sections of the paper, we perform all legal mapping and analysis using CIPA as the primary reference. We do so because it provides stronger protections against wiretapping through its two-party consent requirement and because there is a larger body of precedent-setting case law on its applicability to web tracking.

\subsection{Tech-Law Wiretapping Threat Model}
\label{sec:threat_model}
We consider the following threat model for mapping wiretapping to a website visit by a user:
\begin{itemize}
    \item A user visits a website that includes content from both the website developer and one or more third-party sources.
    \item The website developer controls the primary content served to the user, while embedding third-party resources.
    \item The user interacts with the website using an input device, such as a keyboard.
\end{itemize}
The user interaction with the website is:
\begin{itemize}
    \item Received by the website owner directly from the user via its own code.
    \item Intercepted by a third-party (adversary) which embeds code on the website containing event listeners.
    \item Transmitted to a remote server operated by the third-party. This transmitted information may include sensitive data like passwords or emails.
\end{itemize}
Figure~\ref{fig:input-event-listener} contrasts normal input processing with input processing when a third‑party event listener is installed. In the normal flow (Figure~\ref{fig:input-event-listener-before}), the communication is directed exclusively to the website developer, whereas in the modified flow (Figure~\ref{fig:input-event-listener-after}), the third‑party event listener intercepts the communication \emph{in‑path} and shares it with a remote server.

\begin{figure*}[!htpb]
    \centering
    \includegraphics[width=\linewidth]{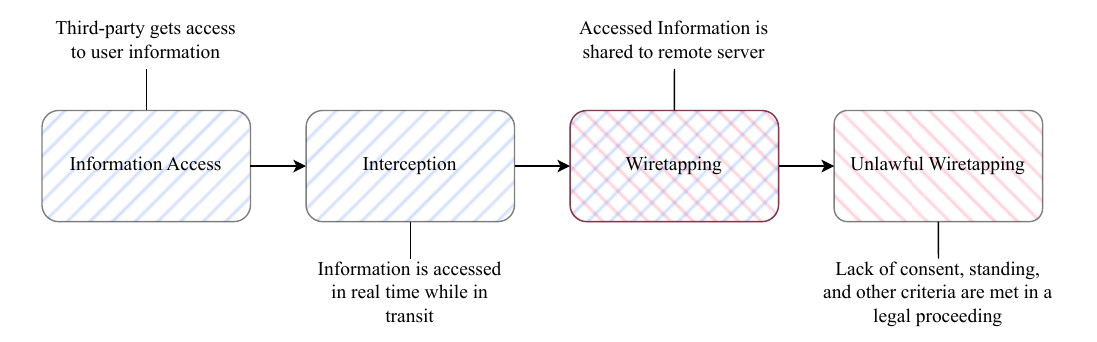}
 \caption{This figure illustrates how terms used in this paper differ based on the mechanisms of information access, the sharing of information, and the legal criteria established by statute and case law.
\protect\tikz[baseline=-0ex]{\protect\fill[
  pattern={Lines[angle=45, distance=3pt, line width=0.8pt]},
  pattern color=blue!20
] (0,0) rectangle (0.6,0.30);} 
indicates that technical methodology is required to measure and assign a term, whereas
\protect\tikz[baseline=-0ex]{\protect\fill[
  pattern={Lines[angle=135, distance=3pt, line width=0.8pt]},
  pattern color=red!30
] (0,0) rectangle (0.6,0.30);} 
indicates that legal understanding and nuance are required to assign a term. In this paper, we focus solely on technical measurement of interception and wiretapping through event listeners.}
    \label{fig:wiretapping-flow-chart}
\end{figure*}

\begin{infobox}
\textbf{\faInfoCircle\ Wiretapping vs.\ Unlawful Wiretapping.} 
Throughout the paper, we use the term \emph{wiretapping} to describe situations where a third party intercepts a user's communication with the webpage and shares it to a remote server.
This use of the term is broader than \emph{unlawful wiretapping}.
For a court to deem wiretapping \emph{unlawful}, additional legal requirements must be assessed under CIPA §631, most centrally whether the interception occurred without the consent of all parties and whether any statutory exceptions apply; suits filed in federal court must also satisfy Article III standing. We do not resolve these questions here.
Figure~\ref{fig:wiretapping-flow-chart} illustrates how we distinguish these terms and the nuances involved in measuring and assigning each term to a scenario in our study.
We defer some of this analysis to Section~\ref{sec:legal_analysis} and Appendix~\ref{app:legal_analysis}.
\end{infobox}

\para{Mapping the Threat Model to \Wiretapping under CIPA.}  
Next, we describe how we map the technical threat model described earlier to the definition of \wiretapping under CIPA.
\textbf{1) Interception of Communications.}  
Because the listener is \emph{in-path}, it captures each keypress or interaction at the moment it is generated, before the data reaches the webpage developer.
Under CIPA §631, listener can be held liable for willfully \emph{reading}, or \emph{attempting to read or learn}, the \emph{contents} of a communication while in transit; however, how that applies to purely on-device logging (without sharing to a remote server) remains unsettled in courts and context-dependent.
This approach primarily targets a very narrow set of tracking scripts that engage in keypress-by-keypress, real-time interception, a practice commonly associated with session-replay or input-capture scripts.

\textbf{2) Sharing of Intercepted Communications.}  
The language of CIPA §631 addresses unauthorized “tapping,” “reading,” or “learning the contents” of communications and does not expressly require disclosure (sharing to a remote server) to establish \wiretapping.
At the same time, district courts have treated off-device sharing of data as a potentially important requirement in certain cases \cite{in_re_tiktok_2024_order,in_re_tiktok_2024}.
To align with this nuance, our threat model includes subsequent sharing of the intercepted information to a remote third-party server.
This ensures that the threat model satisfies the stronger “interception + disclosure” interpretation of CIPA §631.

\textbf{3) Assumed Absence of Consent.}  
For the threat model we assume the user does not interact with any banner, click-through, or other mechanism that would supply affirmative consent to real-time interception.
Thus the communication is treated as non-consensual at the moment of capture.
Further discussion of what constitutes valid consent and how existing case law deals with consent is reserved for Appendix~\ref{app:legal_analysis}.

In summary, our threat model maps the technical elements of \wiretapping under CIPA §631---the interception and disclosure of communications---to the real-time, keypress-by-keypress interception by third-party event listeners.

%% file: parts/measurements.tex
We now describe our methodology to collect and analyze data to detect the use of event listeners and potential wiretappers across a sample of the top-million websites.

\subsection{Website Selection}
From the Tranco list \cite{LePochat.Tranco.2019}, we randomly sampled \nsites websites based on their ranking, dividing them into different buckets for a balanced selection. 
Specifically, we randomly selected 3,000 websites from each of the following rank ranges to ensure a balanced sample across the popularity spectrum: 1–10k, 10,001–50k, 50,001–250k, 250,001–500k, and 500,001–1M. Websites that could not be successfully crawled were excluded from our analysis.
This resulted in a sample totaling \nsites websites.

Prior research has demonstrated that subpages can exhibit different behaviors compared to their corresponding landing pages \cite{Urban.WWW.2020,Demir.Measurements.2022,Aqeel.Sites.2020}. 
To select the final set of webpages for our crawling and analysis, we first visited each randomly selected website's landing page and then gathered 10 subpages (i.e., first-party links) from each, with a preference for those containing HTML forms. 
The focus on forms is due to prior research \cite{Senol2022LeakyformsUSENIX} indicating that privacy violations relevant to our study frequently occur in these contexts, as forms typically request user input. 
To achieve this, we initially crawled over 300k webpages to identify those featuring HTML forms and found, on average, 7 webpages (SD:~4.7; max:~10; min: 0) with forms per website. 
We chose 10 webpages per website, as previous research suggests this number balances the time required for crawling while effectively capturing the website's behavior.
Our final set consists of 125,355 crawled webpages, with 312,741 distinct forms on them.

\subsection{Crawling}
To crawl the webpages and gather the necessary data for our study, we developed a custom web crawler using \textit{Puppeteer} for \textit{Chrome (v134.0)}, a \textit{Node.js} library that automates web browser interactions \cite{puppeteer}. 
We designed our crawler following best practices from related work \cite{Demir.Measurements.2022, Urban.WWW.2020, Senol2022LeakyformsUSENIX, Olejnik.Users.12,Invernizzi.bots.2016}.
We ran all our crawls from within California, USA.

To map our criteria to the webpages, we need to track the data flow (specifically user inputs) on webpages. 
To understand who, when, and how user inputs are accessed, we override all event listeners and invocations to monitor their use in our instrumented browser. We implement all instrumentation in JavaScript and do not modify the browser itself.
We use \texttt{Page.addScriptToEvaluateOnNewDocument()} (via Puppeteer) to make sure our code runs before any page scripts.
We override DOM APIs such as \texttt{addEventListener} and \texttt{removeEventListener} to capture each registration of a handler at the time it is made.
This approach allows us to precisely detect who initializes and invokes event listeners, as well as when and how they are invoked, helping us apply our criteria to the web and determine how a party in the communication accesses and uses user data.

Using our instrumented browser, we begin by crawling the landing pages and then proceed to the identified subpages. 
Our crawler navigates to each webpage and waits for it to fully load, with an additional 3-second buffer to ensure all elements are rendered. 
The crawler then interacts with the webpage by: 1) simulating mouse movements, 2) performing \textit{PageUp}, \textit{PageDown}, \textit{Tab}, \textit{ArrowUp}, and \textit{ArrowDown} keystrokes, 3) filling out all web forms according to the field type \eg inputting an email address for email fields and a phone number for phone fields \textit{without submitting the forms}, as shown in Table \ref{tab:data-shared} 4) entering text into \textit{textarea} fields, and 5) simulating keystrokes in the page body without targeting any specific field, helping us identify if providers are monitoring keystrokes globally, similar to keylogging behavior in a web environment. 
For each interaction, we record the timestamp for further analysis.
Finally, we collect all HTTP requests, responses, cookies, initialization, and invocation details of overridden JavaScript APIs (\eg timestamps, call stack of invocations, parsed data, and code of the invoking functions).

\subsection{Analysis}
\label{subsec:analysis}
To detect data sharing with remote servers, we check if information entered into HTML forms is sent to a remote server.  
Specifically, we look for our input data---whether in raw, encoded, or hashed form---within URLs or request payloads. 

Our analysis uses the following hash and checksum algorithms: MD5, SHA-1, SHA-224, SHA-256, SHA-384, SHA-512, SHA-3 (224, 256, 384, and 512-bit), MurmurHash3 (32-bit, 64-bit, and 128-bit), CRC32, and Adler-32. For encoding and compression, we consider transformations such as Base16, Base32, Base58, Base64, URL encoding, ROT13, HTML entity encoding, and binary string encoding. We also apply compression techniques including Deflate, Gzip, Zlib, Brotli, LZ-string, and LZW, and we systematically test combinations of these transformations, such as compression followed by Base64 or hexadecimal encoding (e.g., Gzip followed by Base64, Deflate followed by hex, Brotli followed by Base64, and others).

Note that our analysis is a lower bound since we can only identify the data leaks if the data we enter in the form is encrypted or decoded with one of these algorithms. 
As related work shows, trackers may use other mechanisms (e.g., salted versions of these algorithms or entirely different methods) for sharing the data~\cite{Senol2022LeakyformsUSENIX}.

%% file: parts/wiretappers.tex
Next, we analyze the data collected from our measurements.
We report the prevalence and usage of event listeners on a sample of top-million websites, and we apply our definition of \wiretapping from Section \ref{sec:threat_model} to identify \wiretappers on a \nsites sample of the top-million websites.
\subsection{Measurement and Prevalence of Event Listeners}
\label{subsec:usage-event-listeners}
We begin by examining the prevalence and characteristics of event listeners across a \nsites-sample of the top-million websites. Our goal is to understand how widely event listeners are used, which entities install them, and what types of interactions they capture. While event listeners are not inherently harmful, some can intercept sensitive user input. In this section, we focus on their overall deployment, laying the groundwork for identifying those with potential privacy implications.

\para{Prevalence.}
The average website has \navgeventlisteners event listeners, and \npcteventlisteners of sites have at least one listener.
\texttt{load} is the most common, followed by \texttt{resize} and \texttt{scroll}, with \texttt{click}, \texttt{keydown}, \texttt{scrollend}, and \texttt{mousemove} also among the top-20.
Table \ref{tab:event-listeners-script-domain-counts} breaks down usage by event type.
\input{tables/event_listeners_script_domain_counts}

\para{Ownership.}
Third-party scripts install \listenersthirdparty of all listeners; while only 18.7\% of first-party\footnote{We differentiate between first- and third-party scripts using Disconnect entity list~\cite{mozilla_disconnect_entitylist}, to avoid under-counting due to use of different hosts and domains.} scripts use event listeners.
Table \ref{tab:percentage-event-listeners-script-domain-sorted} shows the top-20 domains, the share of sites on which they install listeners, their three most common events, and example scripts, organized by script type.
\input{tables/event_listeners_scripts_prevelance_websites_categorized}

\para{Tag Managers.}
Tag managers allow centralized deployment and management of third-party scripts on a webpage.
Google Tag Manager (\textit{googletagmanager.com}) listens to \texttt{scrollend}, \texttt{load}, and \texttt{message} on 67.11\% of sites via \texttt{gtm.js} and \texttt{gtag.js}.
These events allow triggering custom analytics or marketing actions on page load, hide, or scroll depth \cite{GoogleTagManagerScrollDepthTrigger, GoogleTagManagerPageViewTrigger}.

\para{Session Replay.}
Session-replay tools capture user interactions on a webpage for UX analysis.
Microsoft Clarity and Hotjar listen to \texttt{scroll}, \texttt{error}, and \texttt{pagehide} on 10.32\% and 7.26\% of sites, respectively, to measure engagement of users and monitor performance issues on a webpage \cite{HotjarScrollTracking, MicrosoftClarityDataCollection}.

\para{Tracking and Analytics.}
Advertising and analytics scripts track page views and monitor user activity.
Meta’s Pixel (\textit{fbevents.js}, \textit{sdk.js}/\textit{all.js}) uses \texttt{message}, \texttt{pagehide}, and \texttt{pageshow} to capture plugin interactions and view events \cite{MetaFBEventSubscribe, MetaPixelGetStarted}.
%

\para{Miscellaneous.}
Other popular listeners are used by multimedia or security scripts: YouTube’s script (\textit{www-embed-player.js}) captures playback events, and Google (\textit{recaptcha.js}) uses event listeners for bot-detection.

\begin{lesson}
The widespread use of event listeners, especially those set by third-party scripts, shows how deeply embedded external services are in modern websites. While many listeners support functions like scroll tracking, ad visibility detection, or input field monitoring, their scale raises questions about data access and control. These findings suggest that third-party dependencies play a central role in shaping how user interactions are observed, often beyond the website operator's direct oversight.
\end{lesson}

\subsection{Measurement and Prevalence of \Wiretapping}

While our analysis reveals extensive use of event listeners, this does not show \textit{how} intercepted data is used. We therefore perform a targeted examination of those event listeners that intercept communication between user and website to measure extent of \wiretapping across the \nsites sites. To this end, we identify events that directly capture user input---\texttt{keydown}, \texttt{keyup}, and \texttt{keypress}---which \wiretappers can use to intercept users’ on-page communication.

As described in Section \ref{sec:threat_model}, CIPA defines \wiretapping by the interception and disclosure of confidential communication. We therefore label a script as a \wiretapper if it:
\begin{enumerate}
    \item installs an event listener,
    \item performs real-time interception of user-entered communication, and
    \item shares the intercepted data with a third-party server.
\end{enumerate}
\para{Choice of \Wiretapping Event Listeners. }
Specifically, we do not label a script as \wiretapper if it only uses an event listener.
We also do not consider event listeners and APIs which can be used to get access to user communication after the interaction has concluded.
An example of such API is \texttt{document.body}, which can be used to extract the text entered by a user on a website input field.
However, as per our threat model described earlier, the adversary should intercept and then share user communication \emph{in-path} for the script to be considered a \wiretapper.
Table \ref{tab:event-listeners-script-domain-counts-wiretapping} shows the three events we identify as being used for \wiretapping (\texttt{keydown}, \texttt{keyup}, \texttt{keypress}), percentage of websites on which they are used, and percentage of websites where they are used for \wiretapping.
We only focus on these three event types to exercise \emph{utmost} caution in what we classify as \wiretapping, in line with our overall aim in this paper to identify a lower bound on the prevalence of \wiretapping on websites.
\input{tables/event_listeners_types_prevelance_wiretapping}
%

Next, we quantify the prevalence of \wiretapping in the \nsites sample of top-million websites, based on the criteria defined above.
We find that event listeners associated with \wiretapping (i.e., \texttt{keydown}, \texttt{keyup}, \texttt{keypress}) are set on 41.57\% of websites tested, while at least one \wiretapper is identified on \nwtapping of the websites tested.
On average, there are 0.03 \wiretappers per website; on websites with least one \wiretapper, the average is 1.03.
We find \texttt{keydown} \cite{MDNElementKeydownEvent} and \texttt{keyup} \cite{MDNElementKeyupEvent} are the two most commonly used events by \wiretappers, used for \wiretapping on 3.10\% and 1.42\% of websites respectively.
The \texttt{keypress} event, which has been deprecated \cite{MDNElementKeypressEvent}, is still used for \wiretapping on 0.46\% of websites.
These events can be used to directly monitor input of PII (personally identifiable information) of a user such as their email address or phone number entered on a webpage.
Overall, we found that, on 2.09\% of websites, \wiretappers exfiltrated text typed into form fields, and on 1.09\% they exfiltrated email addresses.
It is important to note that all of this information was transmitted to third parties without any form or input submission by the user.
Table~\ref{tab:data-shared} provides a breakdown of the different types of information exfiltrated and the percentage of websites affected.
Later in this section, we show how this PII is collected and misused by \wiretappers.

\input{tables/data_shared_percentage}

\input{tables/wiretappers_types_prevelance_websites}
\begin{lesson}
Our analysis shows that despite our conservative detection criteria, \wiretapping (where user input is captured and shared with third parties) is present on a notable share of websites. \Wiretapping is primarily associated with a narrow set of input events---particularly \texttt{keydown}, \texttt{keyup}, and \texttt{keypress}---which are commonly used to monitor user input.
\end{lesson}

\subsection{Who uses \Wiretapping?}
\label{subsec:wiretapping-use}
Next, we analyze who these \wiretappers are and how \wiretapping is used across the \nsites sample of top-million websites.
As before in Section \ref{subsec:usage-event-listeners}, we divide \wiretappers into four types: tag managers, tracking and analytics, session replay, and miscellaneous.
Table \ref{tab:percentage-wiretappers-script-domain-categorized} shows the categorized list of top \wiretappers identified on \nsites websites.

\para{Tag Managers.} In contrast to the results in Table \ref{tab:percentage-event-listeners-script-domain-sorted}, Google Tag Manager (GTM) is not classified as a \wiretapper on any of the websites tested.
GTM, by default, does not include a trigger related to any of the keyboard related events.
The default triggers used by GTM include page view, click, form submission, history change, custom event, JavaScript error, and timer \cite{Farney2016GoogleAnalyticsTagManager}.
Triggers in GTM are rules or conditions that define when a tag should invoke.
These built-in triggers provide an easy mechanism for monitoring user interactions without requiring the developer to write custom code for each scenario.
Important to note here is that GTM also supports custom event triggers in GTM which allow for greater flexibility, enabling developers to define specific conditions or events that are not covered by the default set.
For example, a custom trigger could be created to activate a tag when a specific button on a webpage is pressed by the user.
However, custom triggers require the webpage developer to implement custom JavaScript logic that listens for and intercepts the desired event.
The information collected is then passed on to GTM, thus bypassing the direct interception of information by GTM \cite{GoogleTagManagerCustomEventTrigger}.

\begin{figure}[!htbp]
    \centering
    \includegraphics[width=\linewidth]{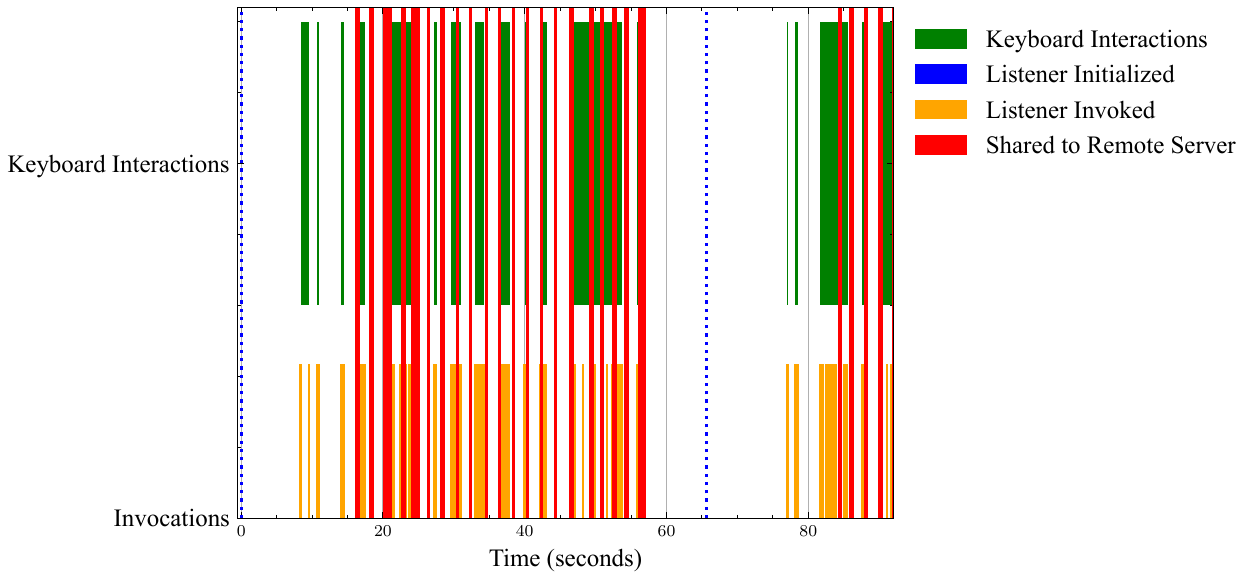}
    \caption{Timeline of keydown events on \textit{vivalavika.com} associated with \textit{yandex.ru} scripts, showing the timeline of user keyboard interactions and the initialization and invocation of event listeners. The x-axis represents time in seconds, while the y-axis categorizes interaction types and invocation events.}
    \label{fig:yandex-invocations}
\end{figure}

\para{Tracking and Analytics.} Cloudflare \textit{cloudflareinsights.com} is the most prevalent analytics \wiretapper which listens to the \texttt{keydown} events on 8.49\% of websites tested, and is found wiretapping on 0.10\% of websites.
Their documentation suggests that Cloudflare prioritizes privacy while providing analytics for a webpage \cite{CloudflareWebAnalytics}.
\textit{yandex.ru} is found listening to \texttt{keydown} and \texttt{keyup} events on 1.75\% of sites, and is found wiretapping on 0.80\% of sites---the most for any domain in our analysis.
Yandex has multiple scripts that \wiretap users, most notable of which is the \textit{tag.js} script used by Yandex's Metrica service \cite{YandexMetrica}.
This service provides analytics services to webpage developers, which include traffic monitoring \cite{YandexMetricaTrafficReport}, session replay \cite{YandexMetricaSessionReplay}, and goal/event monitoring \cite{YandexMetricaGoals}.
Specifically, the session replay functionality of Yandex Metrica monitors user's every interaction with the webpage, resulting in it being classified as a \wiretapper on the highest percentage of sites in our analysis.
Figure \ref{fig:yandex-invocations} illustrates Yandex Metrica monitoring user keyboard interactions and sharing it to its remote server on \textit{valavika.com} website.
We observe near real-time invocation of event listeners set by Yandex---as soon as the keyboard interaction starts on the webpage.
It is noteworthy that this invocation is not limited to typing in fields, but also when keyboard is used to navigate the webpage or random keys are pressed outside of any text fields.
In addition to invocation, we observe periodic sharing of information typed in form fields to remote servers.
As shown in Table \ref{tab:percentage-wiretappers-script-domain-categorized}, Yandex shares both the form text input and the email entered by the user to its remote server.
Typing in non-text fields does not invoke sharing of information for Yandex.

We also observe other analytics and tracking services like Datadog (\textit{datadoghq-browser-agent.com}), Raptive (\textit{adthrive.com)} and Wunderkind (\textit{bounceexchange.com}) among \wiretappers identified in tracking and analytics category.
%
%
Similar to Yandex, Raptive also captures nearly all the information entered by a user on through the keyboard \textit{jaroflemons.com}.
Although it gains access to all the information typed on the webpage, Raptive only shares the emails entered by a user, which is different than Yandex which shares all form inputs.
%

\para{Session Replay.} Our analysis shows that two out of four session replay scripts among the top-40 domains in our analysis are involved in \wiretapping.
%
%
FullStory (\textit{fullstory.com}), which is named as a defendant in multiple lawsuits filed in California (e.g., \cite{Saleh_v_Nike, GrahamvNoom2021}), was found to be involved in \wiretapping on 0.10\% of sites.
In its documentation, FullStory describes in detail the type of information collected by their session replay script \cite{FullStorySessionReplay}.
According to documentation, FullStory attempts to exclude intercepting communication containing sensitive information.
Our analysis shows that FullStory shares text entered in form fields which contains sensitive information such as email and phone number of the user.
%

\begin{figure}[htbp]
    \centering
    \includegraphics[width=\linewidth]{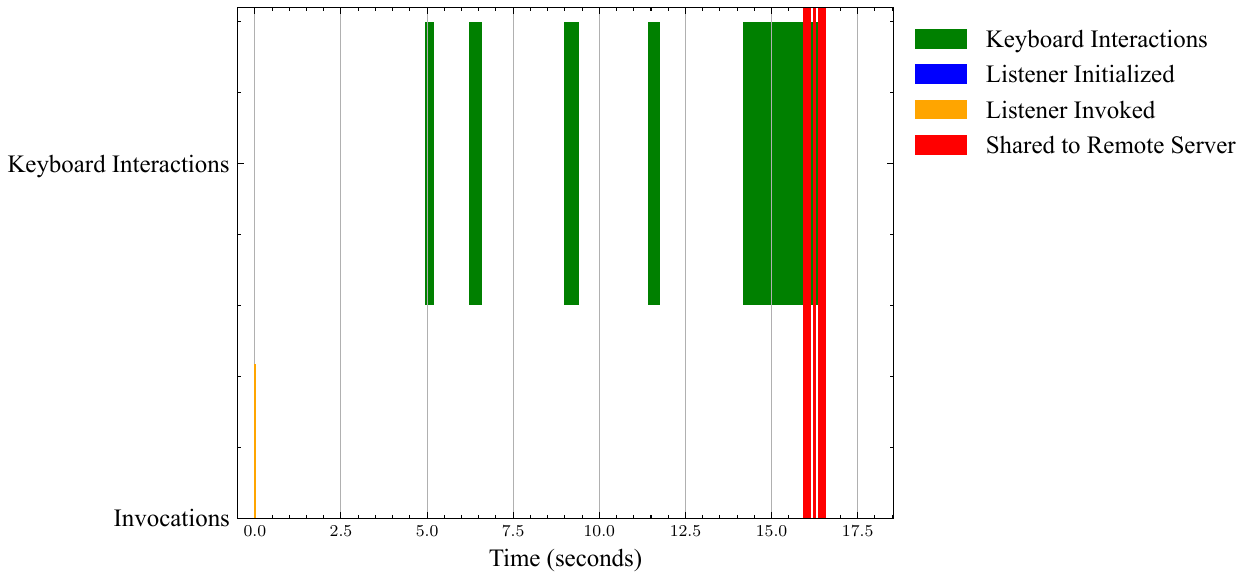}
    \caption{Timeline of keydown events on \textit{vbforums.com} associated with \textit{google.com} scripts, showing the timeline of user keyboard interactions and the initialization and invocation of event listeners. The x-axis represents time in seconds, while the y-axis categorizes interaction types and invocation events.}
    \label{fig:google-invocations}
\end{figure}

\para{Miscellaneous.} In addition to the aforementioned categories, we identify other domains that are involved in \wiretapping.
The most prevalent of these is \textit{google.com}, which is classified as a \wiretapper on 0.6\% of the sites tested due to its \textit{cse\_element\_*.js} script.
This script is used to integrate Google's custom search engine on a webpage \cite{GoogleCustomSearchElement}.
Google uses the keyboard events to monitor user input to the search field and provide relevant search results.
Figure \ref{fig:google-invocations} shows the timeline of Google's custom search engine potentially \wiretapping users on \textit{vbforums.com} website.
Instead of intercepting all keyboard events as is the case with other \wiretappers discussed in this section, Google's event listener is only installed on the custom search field.
Thus, typing on the webpage, using navigation keys, or interacting with other fields does not invoke Google's event listener.
However, as soon as text is typed in the custom search field, the installed event listener is invoked and the information typed is immediately shared to Google's remote server.
This is in stark contrast with the behavior seen for tracking and session replay wiretappers.
This shared information is used by Google for autocomplete and providing relevant search suggestions.

We also observe a number of domains which are used by CDNs to deliver other scripts to webpages.
CloudFront (\textit{cloudfront.net}), identified as \wiretapper on 0.15\% of sites, and jsDelivr (\textit{jsdelivr.net}), which is identified as \wiretapper on 0.06\% of sites, both are used to deliver scripts on a webpage.
Our analysis shows that these scripts are then involved in intercepting and sharing user communication, thus resulting in jsDelivr and CloudFront being identified as \wiretappers.

\para{Known Trackers.} To determine if a \wiretapping script is a known trackers, we make use of filter lists.
We use the \textit{Rust} library \textit{adblock-rust} from \textit{Brave} \cite{Brave_Adblocker_Rust.2024}, utilizing both EasyList  and EasyPrivacy lists to identify tracking requests. 
If any request by a script domain is labeled as tracking, we classify it as a known tracker.
Our analysis shows that \wiretapping is used extensively by known trackers.
Out of all \wiretappers identified in our analysis, 84.60\% of \wiretappers are also known trackers as per EasyList and EasyPrivacy.

\begin{lesson}
We find \wiretapping to be concentrated in a handful of commercial roles:  
(i) analytics and advertising, (ii) session-replay scripts, and
(iii) specialised widgets (e-commerce search, fraud detection, A/B testing).  
We also observe that more than 84 percent of \wiretappers are also known trackers, highlighting \wiretapping's extensive use for web tracking.
\end{lesson}

\subsection{How is \Wiretapped Information Used?}
Next, we investigate the potential use of \wiretapped information.
While it is challenging to make general conclusions about all of the ways in which intercepted information is used at the server-side, we are able to monitor the use of email addresses we entered in form fields as honey-tokens \cite{farooqi2020canarytrap}.
To attribute emails to websites, we entered a unique email address in the form fields on each website. 

We found that email addresses entered into form fields---even without form submission---were intercepted and subsequently used to send unsolicited marketing emails to.
%
Our analysis shows that out of the 50 websites where we observed unsolicited emails being sent to users, we also identify \wiretapping on 3 websites.
On \textit{grabagun.com}, we identified \textit{newrelic.com} as a \wiretapper, which provides session replay services.
On \textit{mongolshop.ru}, we identified \textit{yandex.ru} as a \wiretapper, which is a known tracker (see \ref{subsec:wiretapping-use}).
And lastly on \textit{solo.ind.br}, we identified \textit{vtex.com} as a \wiretapper which provides e-commerce and sales solution to websites \cite{VTEXMarketplace}.

While we were not able to identify \wiretapping on the remaining websites (potentially due to the limitations of our approach described in Section \ref{subsec:analysis}), third-party scripts installed event listeners associated with \wiretapping on 19 of those websites.

\para{Attribution Challenge.}
Once a \wiretapper shares an email address with a third-party marketing platform, the address can potentially be shared, sold, or used by any number of downstream actors.
Subsequent unsolicited messages often originate from domains unrelated to either the website or the original \wiretapper, and the email headers carry no indication of which entity first exfiltrated the address.
Thus, even though our measurement shows that the address \emph{must} have left the page via an in-path interception (we never submitted the form), it becomes practically impossible to identify the specific party that passed the email to the marketer.

\begin{lesson}
We find that keystrokes captured in real-time can trigger unsolicited marketing messages from an opaque network of brokers within hours—even when the user never submits the form.
Subsequent email spam often originates from domains unrelated to the original website or \wiretapper.
\end{lesson}

\subsection{Beyond Event Listeners}
\label{subsec:beyond-wiretapping}
We identify many more instances of user form input collection that does not rely on event listeners.
Specifically, we find evidence of user form input collection on 2,112 (14.5\%) of the websites in our \nsites sample, involving 676 distinct third-party domains.
The top five third parties responsible for collecting user form inputs are \textit{google.com} (present on 275 sites), \textit{hsforms.com} (268 sites), \textit{algolia.net} (170 sites), \textit{google-analytics.com} (167 sites), and \textit{yandex.com} (138 sites).
In these instances, the collection of nearly all user form inputs (excluding password fields) by \textit{google.com}, \textit{google-analytics.com}, and \textit{yandex.com} occurs likely based on how websites integrate these services.
In contrast, \textit{hsforms.com} limits its collection to email addresses.
\textit{Algolia.net}, a search service provider, collects user form inputs from search fields—behavior that aligns more with its provided search functionality.

Instead of event listeners, such user form input collection most commonly occurs through direct DOM access via other interfaces and APIs such as \texttt{document.body} and \texttt{document.querySelector}.
These allow third-party scripts to collect user form inputs without needing to install event listeners.
Although these fall outside our analysis of \wiretapping using event listeners, they illustrate that collection of user form inputs---without submission---by third-party scripts is not uncommon.

\begin{lesson}
Even when keystroke interception does not occur via event listeners, third-party scripts can collect user form inputs using other interfaces and APIs available in web browsers.
Such “silent scraping” of user form inputs occurs on roughly one-in-seven sites we measured. 
\end{lesson}

%% file: tables/event_listeners_script_domain_counts.tex
\begin{table}[!htpb]
    \centering 
\resizebox*{0.7\linewidth}{!}{%
\begin{tabular}{lr}
\toprule
\textbf{Event Type} & \textbf{\% of Websites} \\
\midrule
\texttt{load} & 88.31 \\
\texttt{resize} & 85.94 \\
\texttt{scroll} & 77.54 \\
\texttt{pagehide} & 77.23 \\
\texttt{pageshow} & 76.39 \\
\texttt{message} & 75.03 \\
\texttt{focus} & 74.94 \\
\texttt{blur} & 74.52 \\
\texttt{popstate} & 71.09 \\
\texttt{hashchange} & 67.68 \\
\texttt{test} & 64.95 \\
\texttt{beforeunload} & 59.87 \\
\texttt{click} & 55.13 \\
\texttt{orientationchange} & 54.20 \\
\texttt{keydown} & 51.56 \\
\texttt{visibilitychange} & 51.25 \\
\texttt{scrollend} & 48.64 \\
\texttt{DOMContentLoaded} & 46.01 \\
\texttt{online} & 39.09 \\
\texttt{error} & 38.86 \\
\bottomrule
\end{tabular}
}
\caption{Top 20 most frequently listened to event types and the percentage of websites on which the event listener is installed.}
    \label{tab:event-listeners-script-domain-counts}
\end{table}

%% file: tables/event_listeners_scripts_prevelance_websites_categorized.tex
\begingroup
  \rowcolors{2}{gray!15}{white}
  \begin{table*}[!ht]
    \centering
    \resizebox*{\linewidth}{!}{%
      \begin{tabular}{l c >{\centering\arraybackslash}p{6cm} p{6cm}}
        \toprule
        \textbf{Script Domain} 
          & \textbf{\% of Websites} 
          & \textbf{Event Types} 
          & \textbf{Scripts Used} \\
        \midrule\midrule

        googletagmanager.com      & 67.11 
          & \texttt{message}, \texttt{load}, \texttt{scrollend} 
          & \makecell[l]{gtm.js, gtag.js} \\

        facebook.net              & 30.32 
          & \texttt{message}, \texttt{pagehide}, \texttt{pageshow} 
          & \makecell[l]{sdk.js, fbevents.js, all.js} \\

        gstatic.com               & 28.32 
          & \texttt{message}, \texttt{storage}, \texttt{load} 
          & \makecell[l]{recaptcha\_\_en.js,\\
                        e54*.js,\\
                        recaptcha\_\_ru.js} \\

        googleapis.com            & 20.39 
          & \texttt{message}, \texttt{scroll}, \texttt{load} 
          & \makecell[l]{ima3.js, loader.js,\\
                        jquery.min.js} \\

        doubleclick.net           & 16.17 
          & \texttt{message}, \texttt{scroll}, \texttt{load} 
          & \makecell[l]{pubads\_impl.js,\\
                        gpt.js, rum.js} \\

        googlesyndication.com     & 15.56 
          & \texttt{message}, \texttt{load}, \texttt{scroll} 
          & \makecell[l]{show\_ads\_impl\_fy*.js,\\
                        ufs\_web\_display.js,\\
                        reactive\_library\_fy*.js} \\

        adtrafficquality.google   & 13.97 
          & \texttt{message}, \texttt{load} 
          & \makecell[l]{sodar2.js, VFc2VJAc.js} \\

        youtube.com               & 12.85 
          & \texttt{message}, \texttt{beforeunload}, \texttt{resize} 
          & \makecell[l]{www-embed-player.js,\\
                        base.js, www-widgetapi.js} \\

        licdn.com                 & 10.87 
          & \texttt{ORIBI\_historyChanged}, \texttt{popstate}, \texttt{hashchange} 
          & \makecell[l]{insight.min.js, aj4*.js,\\
                        adq*.js} \\

        clarity.ms                & 10.32 
          & \texttt{scroll}, \texttt{error}, \texttt{pagehide} 
          & \makecell[l]{clarity.js, clarity-extended.js} \\

        2mdn.net                  & 9.58  
          & \texttt{scroll}, \texttt{message}, \texttt{load} 
          & \makecell[l]{client.js, html\_inpage\_rendering\_lib\_*.js,\\
                        express\_html\_inpage\_rendering\_lib\_*.js} \\

        bing.com                  & 9.29  
          & \texttt{scroll}, \texttt{pagehide}, \texttt{pageshow} 
          & \makecell[l]{bat.js, OGe9z2j\_*.br.js,\\
                        fP8*.br.js} \\

        google.com                & 8.63  
          & \texttt{message}, \texttt{load}, \texttt{resize} 
          & \makecell[l]{cse\_element\_\_en.js,\\
                        platform.js, pay.js} \\

        cloudflareinsights.com    & 8.49  
          & \texttt{keydown}, \texttt{pageshow}, \texttt{visibilitychange} 
          & \makecell[l]{beacon.min.js} \\

        cloudfront.net            & 7.47  
          & \texttt{scroll}, \texttt{message}, \texttt{keydown} 
          & \makecell[l]{weddingsNav.*.js, jquery-*.js,\\
                        eyewise-module-consent.js} \\

        criteo.com                & 7.43  
          & \texttt{message}, \texttt{pagehide}, \texttt{visibility-ratio} 
          & \makecell[l]{ld.js, main.e11cabaa.js,\\
                        app.js.*} \\

        hotjar.com                & 7.26  
          & \texttt{scroll}, \texttt{pagehide}, \texttt{error} 
          & \makecell[l]{modules.*.js, *.hash-*.js} \\

        amazon-adsystem.com       & 7.21  
          & \texttt{message}, \texttt{scroll}, \texttt{resize} 
          & \makecell[l]{apstag.js, publisher.js,\\
                        csm\_othersv6.js} \\

        cloudflare.com            & 7.18  
          & \texttt{scroll}, \texttt{message}, \texttt{keydown} 
          & \makecell[l]{jquery.min.js, headroom.min.js,\\
                        lazysizes.min.js} \\

        googletagservices.com     & 7.04  
          & \texttt{message}, \texttt{unload}, \texttt{load} 
          & \makecell[l]{rx\_omid\_video.js, impl\_v106.js,\\
                        impl\_v105.js} \\

        \bottomrule
      \end{tabular}%
    }
    \caption{Percentage of websites on which a script domain sets an event listener, sorted by percentage of sites.}
    \label{tab:percentage-event-listeners-script-domain-sorted}
  \end{table*}
\endgroup

%% file: tables/event_listeners_types_prevelance_wiretapping.tex
\begin{table}[!ht]
    \centering 
  \resizebox*{\linewidth}{!}{    \begin{tabular}{l c r}
\toprule
{} & \textbf{\% of Websites} & \textbf{\% of Websites Involved} \\
\textbf{Event} & \textbf{Event Listener Set} & \textbf{in Potential Wiretapping} \\
\midrule
\texttt{keydown} & 38.89 & 3.10 \\
\texttt{keyup} & 13.25 & 1.42 \\
\texttt{keypress} & 10.46 & 0.46 \\
\bottomrule
\end{tabular}
}
\caption{Events used for wiretapping and the percentage of websites on which the event listener is installed.}
    \label{tab:event-listeners-script-domain-counts-wiretapping}
\end{table}

%% file: tables/data_shared_percentage.tex
\begin{table}[!htpb]
    \centering
    \resizebox*{\linewidth}{!}{%
    \begin{tabular}{l c r}
    \toprule
    \textbf{Data Type} & \textbf{Example Input} & \textbf{\% of Websites} \\
    \midrule
    Form Text & example\_text\_area & 2.09 \\
    Mail & example.email@domain.com & 1.09 \\
    Phone Number & 098765432109 & 0.15 \\
    Password & ExamplePassword1! & 0.01 \\
    URL & example-website.com & 0.01 \\
    \bottomrule
    \end{tabular}}
    \caption{Types of input data shared by wiretappers, with corresponding examples and the percentage of websites on which each type was shared.}
    \label{tab:data-shared}
\end{table}

%% file: tables/wiretappers_types_prevelance_websites.tex
\begingroup
  \rowcolors{2}{gray!15}{white}
\begin{table*}[!htpb]
  \centering
  \resizebox*{0.99\linewidth}{!}{
    \begin{tabular}{l c c c >{\centering\arraybackslash}p{2.3cm} >{\centering\arraybackslash}p{2cm} p{3cm}}
       \toprule
      \textbf{Domain} & \textbf{Known}
                     & \textbf{\% Listeners }
                     & \textbf{\% Identified}
                     & \textbf{Most Used}
                     & \textbf{Data}
                     & \textbf{Scripts} \\
                       &  \textbf{Tracker}                            
                     & \textbf{Installed}
                     & \textbf{Wiretapper}
                     & \textbf{Events}
                     & \textbf{Shared}
                     &  \textbf{Used}                 \\
      \midrule

      \textbf{cloudflareinsights.com}
        & \tikzcmark
        & 8.49
        & \textbf{0.10}
        & \texttt{keydown}
        & Form Text
        & \makecell[l]{vcd*.js, beacon.min.js,\\
                       v84*.js} \\

      \textbf{doubleverify.com}
        & \tikzcmark
        & 5.55
        & \textbf{0.04}
        & \texttt{keypress}
        & Form Text
        & \makecell[l]{dv-measurements*.js,\\
                       dvbm.js} \\

      \textbf{vimeocdn.com}
        & \tikzxmark
        & 3.00
        & \textbf{0.01}
        & \texttt{keydown}
        & Form Text, Mail
        & \makecell[l]{player.module.js, \\ player.js,
                       main-*.js} \\

      ad-score.com
        & \tikzxmark
        & 2.08
        & 0.00
        & \texttt{keypress}
        & —
        & score.min.js \\

      \textbf{yandex.ru}
        & \tikzcmark
        & 1.75
        & \textbf{0.80}
        & \texttt{keydown}, \texttt{keyup}
        & Form Text, Mail, Phone Number, URL
        & \makecell[l]{tag.js, watch.js} \\

      \textbf{google.com}
        & \tikzcmark
        & 1.35
        & \textbf{0.60}
        & \texttt{keydown}, \texttt{keypress}, \texttt{keyup}
        & Form Text
        & \makecell[l]{cse\_element\_\_*.js} \\

      \textbf{userway.org}
        & \tikzxmark
        & 1.23
        & \textbf{0.01}
        & \texttt{keydown}, \texttt{keyup}
        & Form Text
        & \makecell[l]{widget\_app\_*.js, \\ index.js} \\

      parsely.com
        & \tikzxmark
        & 1.07
        & 0.00
        & \texttt{keydown}, \texttt{keyup}
        & —
        & p.js \\

      \textbf{cloudfront.net}
        & \tikzcmark
        & 1.00
        & \textbf{0.15}
        & \texttt{keydown}, \texttt{keyup}, \texttt{keypress}
        & Form Text, Mail, Password, Phone Number
        & \makecell[l]{6981-*.js,\\ reviewsWidget.min.js} \\

      twitter.com
        & \tikzxmark
        & 0.97
        & 0.00
        & \texttt{keydown}
        & —
        & \makecell[l]{embed.*.js, \\ modules-*.js} \\

      brandmetrics.com
        & \tikzxmark
        & 0.93
        & 0.00
        & \texttt{keypress}
        & —
        & 65568.js \\

      \textbf{unpkg.com}
        & \tikzcmark
        & 0.93
        & \textbf{0.01}
        & \texttt{keydown}, \texttt{keyup}
        & Form Text
        & \makecell[l]{web-vitals.*.js,\\ leaflet.js} \\

      \textbf{gstatic.com}
        & \tikzxmark
        & 0.83
        & \textbf{0.03}
        & \texttt{keydown}, \texttt{keyup}
        & Form Text
        & rs*.js \\

      \textbf{adthrive.com}
        & \tikzcmark
        & 0.73
        & \textbf{0.36}
        & \texttt{keyup}, \texttt{keydown}
        & Form Text, Mail
        & adthrive.min.js \\

      \textbf{contentsquare.net}
        & \tikzcmark
        & 0.68
        & \textbf{0.09}
        & \texttt{keydown}
        & Form Text, Mail
        & \makecell[l]{tag.js, 27*.js,\\
                       41*.js} \\

      \textbf{mrf.io}
        & \tikzxmark
        & 0.67
        & \textbf{0.01}
        & \texttt{keydown}
        & Form Text
        & marfeel-sdk.js \\



      \textbf{intercomcdn.com}
        & \tikzcmark
        & 0.61
        & \textbf{0.17}
        & \texttt{keydown}
        & Form Text
        & frame-modern.*.js \\

      heapanalytics.com
        & \tikzxmark
        & 0.61
        & 0.00
        & \texttt{keydown}, \texttt{keyup}, \texttt{keypress}
        & —
        & heap-*.js \\

      \textbf{hsappstatic.net}
        & \tikzxmark
        & 0.60
        & \textbf{0.06}
        & \texttt{keydown}
        & Form Text, Mail
        & embed.js \\

      \textbf{jsdelivr.net}
        & \tikzcmark
        & 0.60
        & \textbf{0.06}
        & \texttt{keydown}, \texttt{keyup}, \texttt{keypress}
        & Form Text, Mail, Phone Number
        & \makecell[l]{tag.js, p-*.entry.js,\\
                       watch.js} \\

      \textbf{fullstory.com}
        & \tikzcmark
        & 0.59
        & \textbf{0.10}
        & \texttt{keydown}, \texttt{keyup}
        & Form Text, Mail, Phone Number
        & fs.js \\

      \textbf{mediavine.com}
        & \tikzcmark
        & 0.57
        & \textbf{0.03}
        & \texttt{keydown}, \texttt{keypress}
        & Form Text, Mail
        & \makecell[l]{wrapper.min.js, \\ 4a*.min.js,
                       13*.min.js} \\

      \bottomrule
    \end{tabular}
  }
  \caption[]{This table shows for each domain: the known tracker status using filter lists, percentage of sites on which it sets wiretapping associated event listener, percentage of sites where it is identified as a \wiretapper, most frequently used events, data shared, and most frequently used scripts. Ticks (\tikzcmark) and crosses (\tikzxmark) indicate true and false.}
  \label{tab:percentage-wiretappers-script-domain-categorized}
\end{table*}
\endgroup

%% file: parts/legal_analysis.tex


In this paper, we present a technical analysis of the extent to which JavaScript event listeners intercept keystrokes and transmit them to third parties, mapping these behaviors to statutory criteria for wiretapping under California’s CIPA §631.
While we situate these findings within relevant legal frameworks, we \textbf{do not} offer legal opinions; nothing herein should be construed as such.

\para{Analytic Scope and Methodological Boundaries.}
\begin{itemize}
  \item \emph{What we flag.} A script is counted if it (i) installs a \texttt{keydown}, \texttt{keyup}, or \texttt{keypress} listener \emph{and} (ii) shares captured inputs to a domain other than the site being visited (i.e., a non-first-party domain). This yields a conservative lower bound on potential \wiretapping.
  \item \emph{Why we ignore consent banners.} CIPA requires consent that is \emph{prior} (before any capture), \emph{specific} (disclosing that keystrokes may be sent to a third party), and \emph{effective} (not a generic notice). Programmatically evaluating these legal requirements at web scale is infeasible, so consent and privacy-policy analysis are left to future work.
  \item \emph{What we omit.} We do not address damages, standing, or monetary harm, nor do we compare CIPA to non-U.S.\ regimes (e.g., GDPR) or to other U.S. state statutes.
\end{itemize}

In summary, courts have recognized that California’s CIPA wiretapping provisions can apply to specific types of website tracking, though interpretations vary and some jurisdictions view similar statutes more narrowly.
The legal landscape around the application of wiretapping to web tracking is still evolving.
More discussion around the legal landscape of wiretapping is reserved for Appendix \ref{app:legal_analysis}.
Our technical analysis can inform ongoing legal debates about whether and how wiretapping laws apply to web tracking. 

%% file: parts/discussion.tex
In this paper, we presented the first tech-law analysis of event listeners based on California's wiretapping statute.
We developed a' threat model that maps the definition of wiretapping in 1967 California Invasion of Privacy Act (CIPA) to the technical operation of web trackers that use event listeners to intercept and share user keystrokes with third parties in real-time. 
We used an instrumented web browser to investigate the use of event listeners for wiretapping on a sample of the top-million websites. 
Our analysis showed that event listeners were installed by third-parties on \nweventlisteners websites to intercept users' keystrokes.
We found that third-party scripts on \textit{at least} \nwtapping websites engaged in wiretapping by further sending the intercepted information to a remote server.
The identified wiretappers commonly include, but are not limited, to known trackers such as session replay scripts.

Our tech-law measurement and analysis presents multiple avenues for future work. 

First, while our work can help identify potential wiretapping for further investigation, it is important to note that our work is limited to a narrow use of event listeners. 
There likely exist other technical mechanisms, such as the Web APIs \cite{MDNDocumentURL} that provide access to the HTML DOM \cite{MDNDOM} and that can be used independently or combined with other types of event listeners (beyond the \texttt{keydown}, \texttt{keyup}, and \texttt{keypress} events considered in Section \ref{sec:wiretapping}), that may also fit the wiretapping criteria.
Future work can consider a broader analysis of such technical mechanisms that may fit the wiretapping criteria.

Second, since consent is a potential defense for websites that have been alleged to engage in wiretapping, an increasing number of websites have started to employ consent banners for California consumers.
Similar to prior work on whether cookie consent banners comply with ePrivacy Directive and GDPR \cite{Degeling.2018, Utz.2019,vanhong.2024.compliance,bielova2024effect,matte2020cookie}, future work can look into whether consent banners that are being geared towards California consumers provide adequate disclosures under CIPA. 
Beyond analysis of disclosures, it is important to investigate whether these consent banners actually work (e.g., web tracking scripts that engage in potential wiretapping are blocked until after consumers consent and that they stop working after consumers withdraw consent).

Third, we performed a limited analysis of the \textbf{\textit{use}} of intercepted information by wiretappers.
We found that user keyboard inputs were intercepted on a variety of form fields and even when users were not typing in a specific form field. 
We were able to detect the use of intercepted email addresses for unsolicited email marketing. 
However, the information intercepted by wiretappers includes not just email addresses but also other types of PII such as phone numbers and free-form text input.
Future work should consider investigating other uses of intercepted information such as for personalized advertising \cite{bashir2016tracing,iqbal2023tracking,liu2024opted}.

Fourth, laws are often written in an abstract, technology-neutral manner so that they can flexibly apply to evolving contexts, a foundational principle of common law \cite{common_law}. 
This abstraction allows courts to reinterpret established legal concepts, such as wiretapping, to new technologies.
Our work contributes here by providing a technical analysis of how web tracking techniques that can align with the wiretapping criteria under CIPA.
As new types of web tracking techniques emerge (e.g. \cite{privacy_sandbox,clearcode_data_clean_room}), there are opportunities to conduct similar work to bridge the gap between emerging technologies and the law.

Finally, beyond the wiretapping provisions in CIPA, other older or established laws (e.g., \cite{law360_vppa_history,pen_register}) are also being reinterpreted and increasingly leveraged for privacy enforcement.
This trend highlights how older or established privacy laws are being reinterpreted and applied to new forms of web tracking.
Future work should consider conducting similar tech-law analysis to apply older or established laws to new forms of web tracking.

%% file: parts/appendix.tex
\section{Ethical Considerations}
\subsection{Guiding Principles and Institutional Review}

This study complies with ethical guidelines for research involving data collection and usage.
Prior to data collection we consulted detailed protocol of our university’s Human Research Protection Program.  
Using the U.S.\ Department of Health and Human Services decision charts for 45 CFR §46\,\cite{HHSDecisionCharts2018}, we determined that the project \emph{does not involve human subjects} because it interacts only with publicly available Web infrastructure and generates exclusively synthetic input.
Accordingly, the work qualified for exemption from further IRB oversight.

\subsection{Data Collection and Minimization}

\begin{itemize}[leftmargin=1.2em]
  \item \textbf{Synthetic traffic only.}  
  Our instrumented Chromium browser visited \nsites public websites and injected \emph{synthetic} keystrokes (e.g., ``example.email@domain.com'') into form fields.  No real user credentials or content were entered or captured.

  \item \textbf{No form submission.}  
  Forms were \emph{not} submitted.  This ensures that the crawler did not alter backend state, trigger transactions, or pollute sites’ analytics beyond the lightweight page-view already incurred by the crawl.

  \item \textbf{PII handling.}  
  The crawler logged only (i) DOM-level metadata indicating which event-listener fired, and (ii) network traces needed to confirm whether the synthetic input left the browser.  

\end{itemize}

\subsection{Potential Risks and Mitigations}

The primary ethical risk is inadvertent collection of non-public user data.
Because the crawler generated the \emph{only} keystrokes captured, no third-party user information could be swept in.
Website owners incurred at most a single page view plus lightweight JavaScript execution—well within normal operational variance.
\section{Open Science}
\label{app:code}
To support reproducibility and future research, we release our measurement code, queries for the full data-processing pipeline, and evaluation scripts. 
All materials, along with supplementary documentation, are available at: \url{https://github.com/javascriptwiretapping/wiretapping-measurement}.
This repository contains the crawler instrumentation, analysis notebooks, and configuration files needed to replicate our study.

\section{Legal Analysis}
\label{app:legal_analysis}
\input{parts/legal_analysis_appendix}

%% file: parts/legal_analysis_appendix.tex
\subsection{Precedential Framework in the U.S.\ Legal System}

The United States follows \emph{stare decisis} (“to stand by things decided”): courts rely on earlier rulings when interpreting the same statute.\footnote{A \emph{statute} is an enactment of a legislative body (e.g.\ Congress or a state legislature), as opposed to judge-made common law or agency regulations.}  
When a prior decision \emph{must} be followed it is called \textbf{binding precedent}; when it may be considered for guidance but is not mandatory it is \textbf{persuasive precedent}.  
The hierarchy---highest to lowest---determines which rulings carry binding force:

\begin{enumerate}[label=(\roman*)]
  \item \textbf{U.S.\ Supreme Court} --- final authority on federal statutes and constitutional questions; its decisions are \emph{binding} on every other court.  
  \item \textbf{U.S.\ Courts of Appeals} (``circuits'') --- their opinions are \emph{binding} on federal district courts \emph{within} that circuit and merely \emph{persuasive} elsewhere.  
  \item \textbf{Federal District Courts} --- decide facts and apply higher-court precedent; their written opinions are \emph{persuasive} (never binding) outside the individual case.\footnote{State courts follow an analogous hierarchy for state statutes such as CIPA, with the state supreme court at the top.}
\end{enumerate}

Because the Electronic Communications Privacy Act (ECPA) and the California Invasion of Privacy Act (CIPA) are statutory, courts begin with the statutory text and, where necessary, consult legislative history and existing precedent.
Because CIPA is a \emph{state} statute, the \textbf{California Supreme Court} is the ultimate authority on its meaning and its rulings bind all California courts.
By contrast, Ninth Circuit interpretations of CIPA are binding only on federal courts within that circuit; they are persuasive, not binding, on California state courts.
In fast-moving technical contexts, disputes typically surface first in the district courts; conflicting outcomes can produce circuit splits that are later resolved by the U.S.\ Supreme Court for federal statutes or by a state supreme court for state statutes.
For these reasons we treat recent lower-court rulings as \emph{guideposts} rather than definitive law.

\subsection{Existing Case Law and Its Precedential Weight}\label{subsec:existing-cases}

We summarize recent decisions that illustrate how wiretap laws are being applied to web technologies.
We state each case’s posture and precedential weight and then extract the narrow takeaway that guides our measurement design.

\para{Interception and sharing of communication (\textit{district court}, persuasive).}
\textit{In re TikTok, Inc.\ In-App Browser Privacy Litigation} (N.D.\ Ill.\ 2024) is a consolidated multi-district litigation (MDL) alleging~\cite{in_re_tiktok_2024,in_re_tiktok_2024_order} that TikTok’s in-app browser injects JavaScript that:

\begin{itemize}
  \item copies every keystroke (including passwords and payment data),  
  \item logs taps and scrolls, and  
  \item forwards that information to TikTok servers for profiling and ad targeting.
\end{itemize}
In an Oct.\ 1, 2024 opinion, the court \emph{largely denied} the motion to dismiss and allowed plaintiffs’ wiretap-based theories to proceed.%
At the pleading stage, the court held only that the complaint alleges facts sufficient to state a \emph{plausible} claim and may move into discovery; the ruling does not resolve the merits or predict the ultimate outcome.%
The court also noted that purely local “keylogger”-style logging may fall outside wiretap statutes, whereas contemporaneous off-device transmission, if proven, could fall within their scope.




\para{Prior consent requirement under CIPA §631 (\textit{court of appeals}, binding in the Ninth Circuit’s federal courts).}
\textit{Javier v.\ Assurance IQ} (9th Cir.\ 2023)~\cite{javier_assurance_2020} is a case against Assurance IQ, which embedded FullStory in a lead-generation webpage.
FullStory began recording mouse movements and keystrokes of a user the moment the page loaded---well before any banner or privacy notice appeared.

The Ninth Circuit, in a reversal of the district court’s initial dismissal, held that CIPA §631(a) requires \emph{prior} consent; consent presented after interception begins is ineffective~\cite{javier_assurance_2020}.
Although, this case was later dismissed by the district court on a statute-of-limitations grounds, the Ninth Circuit’s holding remains binding precedent\cite{Javier_v_ASSURANCE_IQ_LLC}.



\para{SDK recipient liability after trial verdict (\textit{jury verdict in district court}, persuasive; subject to post-trial motions and appeal).}
In \textit{Frasco v.\ Flo Health, Inc.} (N.D.\ Cal.\ 2025), plaintiffs alleged that Flo’s mobile app integrated third-party SDKs that captured in-app health communications and relayed them to Meta’s servers~\cite{Frasco_Class_Cert_Order_2025,Frasco_Verdict_Form_2025,CourthouseNews_Frasco_Verdict_2025,FKKS_Frasco_Takeaways_2025}.
At trial, Flo settled mid-proceeding, and the case continued against the Meta.
A San Francisco federal jury found intentional eavesdropping or recording and absence of consent under CIPA.
Defenses that framed Meta as a mere “service provider” or claimed no downstream “use” of data did not persuade the jury.
This verdict supports treating recipients of off-device transmissions as potential interceptors, regardless of whether or how the data received was used.
This decision, due to being in a district court, may still be refined on appeal.

\para{Jurisdictional divergence on web tracking (\textit{state supreme court}, binding within that state only).}
In \textit{Vita v.\ New England Baptist Hosp.} (Mass.\ SJC 2024), the court held that use of web analytics pixels to monitor page views did not constitute an “interception” under Massachusetts’s wiretap statute; the court invoked the rule of lenity in interpreting a 1960s-era criminal statute in the web context~\cite{Vita_v_New_England_Baptist_Hosp}.
This contrasts with California district-court decisions that have allowed CIPA claims to proceed against similar technologies.

\subsection{Scope and Framing Under CIPA §631}\label{app:legal-scope}

\para{How We use “\Wiretapping.”}
We use \emph{wiretapping} descriptively to denote contemporaneous acquisition of user inputs via event listeners and subsequent transmission to a remote endpoint. Whether such conduct is \emph{unlawful} under CIPA §631 is a separate legal question that turns on statutory elements and defenses and is outside the scope of our technical measurement.

\para{Operational Scope (conservative).}
To keep our analysis narrow and conservative, we:
\begin{itemize}
  \item Flag only scripts that \emph{both} attach key-event listeners (e.g., \texttt{keydown}, \texttt{keyup}, \texttt{keypress}) \emph{and} cause off-device transmission of captured inputs or their derivatives (e.g., hashed or encoded forms). This yields a conservative lower bound on interception and \wiretapping. Notably, alleging interception under §631 does not require proof of post-collection \emph{use} by the recipient.
  \item Do not evaluate \emph{affirmative defenses} such as asserted user consent, “ordinary course of business” processing,\footnote{The “ordinary-course” exception permits a service provider to monitor communications only to the extent necessary to deliver the service itself; it does not authorize unrelated analytics or advertising. See, e.g., \textit{In re Google Inc.\ Gmail Litig.}, 2013 WL 5423918, *11 (N.D.\ Cal.\ Sept.\ 26, 2013).} or lack of harm.
  \item Focus on California’s CIPA §631, which requires consent from \emph{all} parties. We do not interact with consent banners. Under CIPA, consent must be: \textbf{Prior}\footnote{\textit{Javier v.~Assurance IQ, LLC}, 78 F.4th 1312, 1320 (9th Cir.\ 2023) (consent under §631 must be secured before interception).} (obtained before any data capture), \textbf{Specific}\footnote{\textit{In re Google Inc.\ Gmail Litig.}, 2013 WL 5423918, *12 (N.D.\ Cal.\ Sept.\ 26, 2013) (generic privacy notice insufficient at pleading to establish consent).} (the user is informed that inputs may be forwarded to a third party), and \textbf{Effective}\footnote{\textit{Popa v.~Harriet Carter Gifts, Inc.}, 52 F.4th 121 (3d Cir.\ 2022) (interpreting analogous Pennsylvania law and remanding to address adequacy and timing of notice); after remand, district-court proceedings addressed implied consent in light of posted notices.} (a banner that appears after listeners activate, or an opt-out that merely toggles cookies, does not legitimize interception).
  \item \textbf{What we do not decide.} We do not assess the validity of any notice, consent, or statutory exception (e.g., common defenses such as provider exception), nor do we pass any judgment on damages or standing. We also do not analyze elements specific to other provisions (e.g., CIPA §632).
\end{itemize}

\subsection{Rapidly Evolving Doctrine}\label{app:ongoing}

The legal landscape for applying wiretap laws to web tracking is still in flux:

\begin{itemize}
  \item \textbf{Precedent versus outcome.} Appellate rulings remain authoritative even if the case later ends on procedural grounds. For example, \textit{Javier}’s prior-consent rule governs federal courts in the Ninth Circuit notwithstanding later dismissal on timeliness.
  \item \textbf{Active docket and splits.} District courts disagree on whether embedded analytics or session replay vendors act as third-party “interceptors” or as tools of the first party.
  \item \textbf{State variation.} Outside California, state courts may read their own wiretap statutes differently. \textit{Vita} narrows Massachusetts liability for website tracking relative to California decisions applying CIPA.
\end{itemize}

Accordingly, readers should treat current case law as provisional; new appellate or legislative developments could significantly alter the framework for assessing JavaScript‐based interception.  

\FloatBarrier
\begin{table*}[!htb]
\centering
\small
\begin{tabular}{p{3.7cm}p{2.5cm}p{2.3cm}p{2cm}p{6cm}}
\toprule
\textbf{Case} & \textbf{Court \& Year} & \textbf{Status} & \textbf{Binding} & \textbf{Takeaway} \\
\midrule
\textit{Javier v.\ Assurance IQ}~\cite{javier_assurance_2020,Javier_v_ASSURANCE_IQ_LLC} & 9th Cir.\ 2022/2023 & Dismissed on statue-of-limitations grounds & Ninth Circuit federal courts & Consent must be prior to interception; banners after listeners fire are ineffective under §631(a). \\
\textit{In re TikTok In-App Browser}~\cite{in_re_tiktok_2024,in_re_tiktok_2024_order} & N.D.\ Ill.\ 2024 & In discovery phase & No (persuasive) & JavaScript-based interception and off-device transmission to satisfy CIPA §631 can survive in courts \\
\textit{Frasco v.\ Flo Health}~\cite{Frasco_Class_Cert_Order_2025,Frasco_Verdict_Form_2025,CourthouseNews_Frasco_Verdict_2025,FKKS_Frasco_Takeaways_2025} & N.D.\ Cal.\ 2025 & Jury verdict & No (persuasive) & SDK recipient can be treated as an interceptor where off-device transmissions occur; lack of proof for downstream “use” does not absolve wiretapping claims. \\
\textit{Vita v.\ New England Baptist Hosp.}~\cite{Vita_v_New_England_Baptist_Hosp} & Mass.\ SJC 2024 & State high court & Binding only in Mass. & Pixel-based page-view tracking not an interception under the Massachusetts statute; illustrates jurisdictional divergence from CIPA cases. \\
\bottomrule
\end{tabular}
\caption{Selected decisions and narrow takeaways relevant to web based wiretapping.}
\end{table*}